\documentclass[12pt]{article}
\usepackage{latexsym,graphicx}
\newcommand{\be}{\begin{equation}}
\newcommand{\ee}{\end{equation}}
\usepackage[left=2.5cm,top=2.5cm,right=2.5cm,bottom=1.5cm]{geometry}
\usepackage{float}
\usepackage{epstopdf}
\usepackage{cite}
\begin{document}
\begin{center}
\large{\bf{Non-exotic wormholes in 4-D Einstein-Gauss-Bonnet gravity}}\\
\vspace{10mm}
\normalsize{ Ambuj Kumar Mishra$^1$, Shweta$^2$,  Umesh Kumar Sharma$^3$,}  \\
\vspace{5mm}
\normalsize{$^{1,2,3}$Department of Mathematics, Institute of Applied Sciences and Humanities, GLA University,
	Mathura-281 406, Uttar Pradesh, India}\\

\vspace{2mm}
$^1$E-mail:  ambuj\_math@rediffmail.com \\
\vspace{2mm}
$^2$E-mail: shwetaibs84@gmail.com  \\
 \vspace{2mm}
$^3$E-mail: sharma.umesh@gla.ac.in  \\
\vspace{2mm}
\end{center}
\vspace{10mm}
%\date{}
%\maketitle
\begin{abstract}
In the present paper, we investigate wormholes in 4D-Einstein-Gauss-Bonnet gravity without the requirement of exotic matters. We have taken the radial dependent red-shift function  $\phi=\ln  \left( {\frac {r_{{0}}}{r}}+1 \right)$ and shape function $b(r)={\frac{r_{0}\ln(r+1)}{\ln({r_0}+1)}}$ as well as anisotropic matter sources through equation of state (EoS) $p_r(r) = \omega \rho(r)$. We examine the energy conditions, flaring out condition, throat condition and anisotropic parameter. The volume integral quantifier is also analyzed to validate the existence and stability of the WH solution. We find, for -ve branch wormhole solutions satisfy the null energy conditions (NEC) throughout the entire space time and +ve branch reduces exactly to Morris-Thorne of GR.

\end{abstract}



%%%%%%%%%%%%%%%%%%%%%%%%%%%%%%%%%%%%%%%%%%%%%%%%%%%% Section 1 %%%%%%%%%%%%%%%%%%%%%%%%%%%%%%%%%%%%%%%%%%%%%
\section{Introduction}

Within the framework of general relativity, the wormholes are characterized as the hypothetical tunnel shaped structures or bridges, providing a connection between two asymptotically flat regions or infinities in space-time. If we go through the available literature we find that the concept of wormhole was first introduced by Flamm \cite{ref1} who suggested a tunnel like structure in the Schwarzschild solution. The most simple example of wormhole space time is the maximally extended Schwarschild spacetime. Einstein and Rosen \cite{ref2} put forward the concept of static wormholes connecting two exterior points of a Schwarzschild black hole space-time in 1935 by transformation of radial coordinates. These static wormholes are now referred to as the Einstein-Rosen bridge. The Einstein-Rosen bridge can be indicated as a static wormhole without a throat. Fuller and Wheeler \cite{ref3} suggested that a coordinate system can be introduced to constitute the Einstein-Rosen bridge with a throat. They first used the name wormhole for these bridges for microscopic charge carrying wormholes. They denied the idea of stable wormholes by suggesting that it would crumple as soon as it forms. Also if such a wormhole unlatches it would compress too fast to pass even a single photon though its throat.

The Schwarzschild solutions are dynamic non traversable wormholes satisfying energy conditions that can also be seen in Penrose diagram, deriving from the fact that this space time region has no wormhole throat of null hyper-surface \cite{ref4,ref5,ref6}. Morris and Thorne gave the static, spherically symmetric traversable wormhole connecting two asymptotically flat regions in space-time \cite{ref7} which subsequently was shown to be in sync with tachyonic massless scalar field\cite{ref8,ref9}. Later Morris and Thorne along with Yurtsever explored the possible existence of traversable wormholes allowing the matter and radiation to travel through it and explored the idea of making a time machine \cite{ref10,ref11,ref12}. Researchers these days are more interested in these traversable Lorentzian wormholes lacking any horizons or singularities.\cite{ref7,ref13}.

In GR, for static wormholes to exist, the null energy conditions must be violated, which points to the presence of exotic matter or ghost energy in the wormhole throat which possess negative energy (exotic properties)\cite{ref4,ref14,ref15,ref16,ref17,ref18,ref19}. In the case of two asymptotically flat regions it is not possible to travel between spatial  regions due to this violation\cite{ref21,ref22}. The construction of WH satisfying WEC can be possible in classical general relativity\cite{ref23,ref24, refa}.

Recently, intensive efforts have been made to find the WH in the modified as well as higher dimensional theories of gravity \cite{ref25,ref26,ref27,ref28,ref29,ref30,ref31,ref32,ref33}
The prime objective of these theories is to avoid the existence of phantom fluid in the throat\cite{ref16}. It has been proved that the energy conditions may be respected by the dynamical spherically symmetric WH \cite{ref24,ref34,ref35,ref36}. A thin shell wormhole has also been produced with reduced use of exotic or nonstandard matter  by limiting it only to the throat \cite{ref37,ref38,ref39,ref40,ref41,ref42,ref43,ref44}. In higher dimensional theories and modified theories of gravity with higher order curvature, the WH may respect the energy conditions\cite{refa,ref25, ref b,ref45,ref46,ref47,ref48,ref49}. Modified gravity theories have indicated that it is the higher order curvature term and not the exotic matter that contributes to the gravitational fluid and it is the ordinary matter in the throat which is derived from the non violation of the energy conditions.

The most general gravitational theory in n-dimension is the Lovelock gravity\cite{ref50} in which the n-dimensional Lorentzian WH geometries have been explored\cite{ref51}. In Lovelock gravity the throat radius of WH depends on the Lovelock coefficients and shape function, also on the dimensionality of space-time. The region of ordinary fluid around the throat is increased by the higher order Lovelock terms having negative coupling constants. The n-D Einstein-Gauss-Bonnet theory or precisely the n-D EGB theory is the second order lovelock gravity\cite{ref52} which has evolved as a significant theory in which Lorentzian WH solutions have been investigated. The lovelock polynomial depicts the second order quasilinear field equations obtained from the contractions on Riemann curvature tensor. It is seen that in most of the cases the throat radius depends upon the dimensionality n of space time and the GB coupling constant $\alpha $. A full-fledged study to analyze the n-dimensional static WH solutions is done in EGB gravity with and without cosmological constant. The study further is generalized with the assumption that space time has symmetries corresponding to the isometries of an $\left( n-2 \right) $ dimensional maximally symmetric space having the finite sectional curvature\cite{ref53}. It was assumed that the metric must be at least $ C^2 $ and the subspace must be compact.

EGB is the generalization of GR to the higher dimensions given by Lanczos\cite{ref54} which can be established in the low energy limit of string theory\cite{ref55,ref56,ref57}. It is supposed to be stable when expanded about flat space time deriving the non exotic gravitational self interactions\cite{ref58,ref59}. The EGB theory does not participate in the gravitational dynamics as it is topological in 4-D due to the Lagrangian being a total derivative. Hence to achieve the non trivial gravitational dynamics D must be greater than 4  \cite{ref60}.  Glavan and Linn reconstructed the GB coupling constant $\alpha$ as $\frac{\alpha}{\left(D - 4 \right)}$ and taking the limit as $D \rightarrow 4$  in the field equations, the non trivial dynamics in 4-D is achieved. Using 4-D EGB theory Glavan and Linn suggested static, spherically symmetric vacuum black holes that are free from singularities \cite{ref60,ref61,ref62}. Extensive work in the 4-D EGB theory has been done in order to find the solutions for  black holes having distinct properties \cite{ref63,ref64,ref65,ref66,ref67,ref68,ref69}.  A text related to the non acceptance of 4-D EGB theory is also produced by Gurses and Sisman \cite{ref70}. The higher order asymptotically flat wormhole solutions are examined and the specific solutions obeying the weak energy condition throughout the space time are achieved in the backdrop of GB gravity\cite{refa}. The effects of GB term are also studied taking a particular radial dependent redshift function and applying a certain equation of state. Kanti P. investigated the properties of Lorentzian WH in the dilatonic 4-D EGB theory\cite{ref71}. He suggested that the exotic matter is not required for the existence of such WH and some of them are linearly stable.

Recently, Mehdizadeh \cite{ref71a} investigated dynamical wormholes in Einstein-Gauss-Bonnet gravity by using specific shape function and scale factor $R(t)$. He obtained wormhole solutions that obey the weak energy conditions. Jusuffi et. al \cite{ref13} obtained wormholes in 4D EGB gravity for coupling constant $\alpha \geq 0$. They considered a constant redshift function i.e. $\phi'(r)=0$ and different shape functions in isotropic and anisotropic source material and found that in general null energy conditions (NEC) are violated. In the present work, we are motivated to find non-exotic wormholes in 4D-EGB gravity i.e. wormhole solutions that satisfy null energy conditions throughout the space-time. This paper is structured as follows: In Sec. 2, we show  a brief review of 4D-EGB gravity. In Sec. 3 we discuss the wormhole metric and solution of field equations in 4D-EGB gravity. In Sec. 4, exact wormhole solutions are analyzed in three different cases. The volume integral quantifier (VIQ) is discussed in Sec. 5. Concluding remarks are given in Sec. 6.  

%%%%%%%%%%%%%%%%%%%%%%%%%%%%%%%%%%%%%%%%%%%%%%%%%%%% Section 2 %%%%%%%%%%%%%%%%%%%%%%%%%%%%%%%%%%%%%%%%%%%%%%%%
\section{The EGB Model} 

The action that determines the D-dimensional EGB gravitational theory is formulated as

\begin{equation}\label{eq1}
S = \frac{1}{16 \pi} \int d^D x\sqrt{-g} \left[ R + {\frac{\alpha}{D - 4}} \mathcal{L}_{GB} \right] + {S}_{matter}  ,  
\end{equation}
In this equation $\alpha$ is the coefficient of Gauss Bonnet coupling having dimension $\left(length\right) ^2$  whereas $g$ gives the determinant of the metric ${g}_{\eta \zeta}$. The matter field appearing in theory is given by ${S}_{matter}$. Also the Lagrangian $\mathcal{L}{GB}$ is elucidated as 

\begin{equation}\label{eq2}
\mathcal{L}_{GB} =  {R}^{\eta \zeta \chi \psi} {R}_{\eta \zeta \chi \psi} - 4  {R}^{\eta \zeta}  {R}_{\eta \zeta} +R^2,  
\end{equation}
where ${R}_{\eta \zeta \chi \psi}$ and ${r}_{\eta\zeta}$gives the  Reimannian tensor and Ricci tensor  respectively. Ricci scalar is symbolized as R.\\
The field equations for the theory are obtained when the  action (\ref{eq1}) is varied w.r.t. ${g}_{\eta \zeta}$, which are specified as \cite{ref72}

\begin{equation}\label{eq3}
{G}_{\eta \zeta} + \left[ {\frac{\alpha}{D - 4}} \right] {H}_{\eta \zeta} =  {8 \pi} {T}_{\eta \zeta},  
\end{equation}
here, ${T}_{\eta \zeta} $ is the energy momentum tensor and ${G}_{\eta \zeta}$ is Einstein tensor that can be expressed as

\begin{equation}\label{eq4}
{G}_{\eta \zeta} = {R}_{\eta\zeta} + \frac{1}{2} R {g}_{\eta\zeta},  
\end{equation}
Also, 

\begin{equation}\label{eq5}
{H}_{\eta \zeta} = 2 \left[R {R}_{\eta\zeta} - 2 {R}_{\eta\psi}{R}^{\psi}_{\zeta} - 2 {R}_{\eta\psi \zeta\chi}{R}^{\psi\chi} -  {R}_{\eta\zeta\chi\delta}{R}^{\psi\chi\delta}_{\zeta} \right] -  \frac{1}{2}\mathcal{L}_{GB} {g}_{\eta\zeta},  
\end{equation}
which is the Lancoz tensor.

The Novel 4-D EGB theory  was introduced to avoid the 4-D Lovelock theorem by rescaling the GB coupling coefficient $\alpha$ as $ \frac{\alpha}{\left[D-4 \right]} $ and having the limit as $D \rightarrow 4$ so that significant effects can be measured on local gravity \cite{ref60}. Although this theory does not possess standard field equations in 4-D as the GB term is a total derivative in space-time. But infusing rescaled-coupling constant $ \frac{\alpha}{\left[D-4 \right]}$ with maximally symmetric space time having curvature scale K \cite{ref59}, one can get field equations without letting the variation of action to disappear, as 

\begin{equation}\label{eq6}
\frac{{g}_{\eta \psi}}{\sqrt{-g}} \frac{\delta \mathcal{L}_{GB}}{\delta {g}_{\zeta \psi}} = \frac{ \alpha \left(D-2 \right) \left(D-3\right)}{2\left(D-1 \right)} K^2 {\delta}^{\zeta}_{\eta},  
\end{equation}
Previously, solutions for Wormhole in 4-D EGB theory are obtained by using various regularization methods getting exactly same 4D spherical solutions \cite{ref13,ref62,ref72,ref73,ref74,ref75,ref76}.

%%%%%%%%%%%%%%%%%%%%%%%%%%%%%%%%%%%%%%%%%%%%%%%%%%%%%%%%%  done below
\section{Wormhole Metric 4-D EGB gravity and solution of field equation}

For wormhole in 4D-EGB gravity we consider the general spherically symmetric, static D-dimensional metric written as \cite{ref42} 

\begin{equation}\label{eq7}
dz^2=-{\exp}^{2 \phi (r)} dt^2 + \frac{dr^2}{1- {\frac{b(r)}{r}}} + {r^2} d\Omega ^2 _ {D-2}
\end{equation}

\begin{equation}\label{eq8}
d\Omega ^2 _  {D-2} = d\theta _1 ^2  + \Sigma  _ {i = 2}^{D-2} \prod_{j=1}^{i - 1} {\sin^2}{\theta_j} d\theta_i ^2 
\end{equation}
The redshift function of an in-cursive object is given by $\phi(r)$ which must be finite and must not disappear at the throat to avoid the presence of any singularity around the wormhole. In this present study we are going to take two cases for the red shift function. One by taking a constant redshift function and another case is studied by taking a variable redshift function.

Here the shape function $b(r)$ is the one which is responsible to define the shape of the wormhole,  hence it has to comply with certain conditions to support the wormhole geometry. The throat condition $0 < 1-\frac{b(r)}{r}$ for   $r>r_0$ is satisfied by function $b(r)$ along with the condition $b(r_0)= r_0$.  The minimum value of $r$ is $r_0$, which is known as throat radius and  $r_0 \leq r \leq \infty$. The condition $b'(r_0) <1$ is called a flaring out condition that signifies the minimum radius of the throat and the shape function has to obey this condition within the throat to ensure the traversability through the wormhole. Asymptotically flat space- time- geometry is achieved when $\frac{b(r)}{r}\rightarrow 0  \qquad as \qquad |r|\rightarrow \infty$ which also is an essential condition for the eligible shape function.

The stress-energy-momentum tensor for the anisotropic fluid as the matter filled in throat is defined as 

\begin{equation}\label{eq9}
T_i ^ \zeta = \left(\rho(r) + p_l(r) \right ) u^\zeta u _i + p_l(r) g^\zeta _i + \left(p_r(r) - P_l(r)\right) \beta_i \beta^\zeta
\end{equation}
where $\rho(r)$, $p_l(r)$ and  $p_r(r)$  are energy density, tangential pressure and radial pressure respectively, whereas $ u^\zeta$ is the 4-velocity in 4-D EGB theory and $\beta^\zeta$ represents the space-like vector in the direction parallel to radius. By infusing the tensor (\ref{eq9}) in the metric (\ref{eq7}) and taking the limit as $D \rightarrow 4$, we find the components $\rho(r) $, $p_r(r)$ and $p_l(r)$ as

\begin{equation}\label{eq10}
\rho(r)  = {\frac{\alpha b(r)}{8 \pi r^6} }{\left ( 2 r b'(r) - 3b(r)\right)} +{\frac{b'(r)}{8\pi r^2}}.
\end{equation} 

 \begin{equation}\label{eq11}
 p_r(r) = \frac{\alpha b(r)}{8 \pi r^6} \left[ 4 \left(r - b(r)\right) \phi '(r) + b(r)\right] -\frac{2 \left(r - b(r)\right)\phi '(r)}{8 \pi r^2} - \frac{b(r)}{8 \pi r^3}.
\end{equation} 
 		
\begin{eqnarray}\label{eq12}
p_l(r) &=& \frac{r - b(r)}{8r\pi}\left[\left(\phi^{''}(r) + \phi^{'2}(r)\right) \left(1+\frac{4\alpha b(r)}{r^3} \right) + \frac{1}{r}\left(\phi^{'}(r)- \frac{r b^{'}(r) - b(r)}{2r \left(r - b(r)\right)}\right) \left(1 - \frac{2 \alpha b(r)}{r^3}\right)\right.\nonumber\\
& -&
\left.\left(\frac{\left(r b^{'}(r) - b(r)\right)\phi^{'}(r)}{2r \left(r - b(r)\right)}\right) \left( 1 - \frac{8 \alpha}{r^2} + \frac{12 \alpha b(r)}{r^3} \right)\right] + \frac{2\alpha b^2(r)}{8\pi r^6} 
 \end{eqnarray}

%%%%%%%%%%%%%%%%%%%%%%%%%%%%%%%%%%%%%%%%%%%%%%%%%%%%%%%%%%%%%%%%%%%%%%%%%%%

%%%%%%%%%%%%%%%%%%%%%%%%%%%%%%%%%%%%%%%%%%%%%%%%%%%%%%%%%
\subsection{Energy conditions}
The stability and existence of the traversable wormhole lies upon some inequalities in terms of energy density, radial pressure and the tangential pressure related to the energy momentum tensor. These inequalities, generally called energy conditions, play an important role in finding the solutions of wormholes. The null energy condition or NEC is generally violated by the matter filled in the wormhole which indicates the negative pressure or repulsive nature of the matter. The roots of NEC are found in Raychaudhuri equation, which is 
\begin{equation}\label{13}
\frac{d\theta}{d\tau} + \frac{1}{2}\theta^{2}+\sigma_{\eta \zeta} \sigma^{\eta \zeta}-\omega_{\eta \zeta}\omega^{\eta \zeta}= -R_{\eta \zeta}k^{\eta}k^{\zeta}.
\end{equation}
Here $k^{\eta}$ is the vector field.The Shear ,a spatial tensor is given by $-R_{\eta\zeta}k^{\eta}k^{\zeta}$ with $\sigma^{2}= \sigma_{\eta\zeta}\sigma^{\eta\zeta} \geq 0$ and $\omega_{\eta\zeta}\equiv0$

The positivity condition  $R_{\eta\zeta}k^{\eta}k^{\zeta}\geq0$ in the Raychaudhuri equation gives the regularized NEC, $T_{\eta\zeta}^{eff}k^{\eta}k^{\zeta}\geq0$,
via field equation of the modified gravity\cite{ref49}. Here $T_{\eta\zeta}^{eff}$ is the effective stress energy tensor. The NEC $T_{\eta\zeta} k^{\eta}k^{\zeta} \geq 0\Leftrightarrow NEC$,also written as  $\forall i, \rho(r) + p_{i}\geq 0 \Leftrightarrow NEC$ in terms of principle pressures indicates the non-negativity of the pressure as observed by an observer. The condition $T_{\eta\zeta} k^{\eta}k^{\zeta} \geq 0\Leftrightarrow WEC$ is known as the weak energy condition or WEC for a time like vector. As functions of principle pressures this condition is given as  $\rho(r)\geq 0$ and $\forall i, \rho(r) + p_{i}\geq 0 \Leftrightarrow WEC $. It simply signifies that the energy density must be positive locally,for a time like object. The violation of strong energy condition or SEC  $(T_{\eta\zeta}-\frac{T}{2}g_{\eta\zeta}) k^{\eta}k^{\zeta} \geq 0\Leftrightarrow SEC$, where $T$ denotes trace of energy momentum tensor is must to have the wormhole geometry.The violation of SEC describes the inflation of universe.  In terms of principal pressures SEC is given  as $T = -\rho(r) + \sum_{j} p_{j}$ and  $\forall j, \rho(r) + p_{j}\geq 0,  \rho(r) +\sum_{j} p_{j}\geq 0  \Leftrightarrow SEC$.
The dominant energy condition  $DEC \Leftrightarrow T_{\eta\zeta} k^{\eta}k^{\zeta} \geq 0$ where $T_{\eta\zeta}k^{\eta}$ is not space-like $\rho(r) \geq 0$ and $\forall i, p_{i}\in \left[-\rho(r), +\rho(r)\right] \Leftrightarrow DEC$ restricts the flow of energy and mass to the speed of light i.e. the rate of flow can not be more than the speed of light. It also suggests that energy density is positive locally.

%%%%%%%%%%%%%%%%%%%%%%%%%%%%%%%%%%%%%%%%%%%%%%%%%%%%%%%% Section 4 %%%%%%%%%%%%%%%%%%
\section{Solution of the Wormhole Model} 
Here we have taken the logarithmic shape function    $b(r)={\frac{r_{0}\ln(r+1)}{\ln({r_0}+1)}}$, where $r_0$ represent the throat radius \cite{ref76a, ref76b}. Further redshift function $\phi(r)$ should be finite and non-zero to avoid the existence of horizons. So it may be constant or variable. To justify asymptotically flatness of wormhole here we have taken variable redshift function $\phi(r)=\ln  \left( {\frac {r_{{0}}}{r}}+1 \right)$ (see in \cite{ref76c}).  We examine for the solution of wormholes and 
the EoS function giving the relation between energy density $\rho(r)$ and radial pressure $p_r(r)$ is taken as 
 
 \begin{equation}\label{eq14}
p_r(r) = \omega \rho(r),
\end{equation}
Here, the EoS parameter $\omega$ or radial state parameter, as is called in terms of radial pressure, defines the characteristic of the fluid filled in wormhole throat. For astrophysical as well as cosmological phenomena, different EoS are used by theoretical physicists. The equation of state of fluid backing wormhole
theory has an important role in investigating wormhole geometry.
A well-known possibility for cosmic fluids with  EoS is vacuum energy $\omega=-1$. Mathematically, the vacuum energy is comparable to a CC (cosmological constant). The
matter sources with $ -1<\omega<-1/3$ are referred to as quintessence dark energy and accelerate the universe. Phantom: a hypothetical exotic matter that has EoS $\omega<-1$, also  accelerates the universe and is commonly used as a principal applicant for dark energy. The phantom fluid is observed
to be viable with the classical tests of cosmology 	like cosmic microwave background (CMB), mass power spectrum and anisotropy. 

Taking the above shape $b(r)$ and specific redshift function $\phi(r)$  with the help of (\ref{eq10}), (\ref{eq11}) and (\ref{eq12}), the energy density $\rho(r)$, radial pressure $p_r(r)$ and tangential pressure $p_l(r)$ are obtained as  

\begin{equation}\label{eq16}
\rho(r) = {\frac {r_{{0}} \left( -3\,\alpha\,r_{{0}} \left( r+1 \right) 
		\left( \ln  \left( r+1 \right)  \right) ^{2}+2\,\alpha\,r_{{0}}\ln 
		\left( r+1 \right) r+\ln  \left( r_{{0}}+1 \right) {r}^{4} \right) }{ 8
		\left( r+1 \right)  \left( \ln  \left( r_{{0}}+1 \right)  \right) ^{2
		}{r}^{6}\pi }}
\end{equation}  

\begin{equation}\label{eq17}
p_r(r) = -{\frac {r_{{0}} \left( 2\, \left( \ln  \left( r_{{0}}+1 \right) 
	\right) ^{2}{r}^{4}+r\ln  \left( r+1 \right)  \left( 4\,\alpha\,r_{{0
	}}-{r}^{2}r_{{0}}+{r}^{3} \right) \ln  \left( r_{{0}}+1 \right) -
	\alpha\,r_{{0}} \left( \ln  \left( r+1 \right)  \right) ^{2} \left( 5
	\,r_{{0}}+r \right)  \right) }{8 {r}^{6} \left( \ln  \left( r_{{0}}+1
	\right)  \right) ^{2} \left( r_{{0}}+r \right) \pi }}
\end{equation}

\begin{eqnarray}\label{eq18}
p_l(r) &=& \frac{1}{16}\,r_{{0}} \left( -6\,r_{{0}}\alpha\, \left( r+{\frac {19}{3}}\,r_{
	{0}} \right)  \left( r+1 \right)  \left( \ln  \left( r+1 \right) 
\right) ^{2}+r \left(  \left( r+1 \right)  \left( {r}^{3}-2\,{r}^{2}r
_{{0}}+28\,\alpha\,r_{{0}} \right) \ln  \left( r_{{0}}+1 \right)\right.\right.\nonumber\\
 &+&
\left.\left.2\,\alpha\,r_{{0}} \left( 7\,r_{{0}}+r \right)  \right) \ln  \left( r+1\right) +2\, \left(  \left( r+1 \right) \ln  \left( r_{{0}}+1
\right) {r}^{2}-1/2\,{r}^{3}-4\,\alpha\,r_{{0}} \right) \ln  \left( r
_{{0}}+1 \right) {r}^{2} \right)\nonumber\\
&\times&
 {r}^{-6} \left( \ln  \left( r_{{0}}+1
\right)  \right) ^{-2} \left( r_{{0}}+r \right) ^{-1} \left( r+1
\right) ^{-1}{\pi }^{-1}
\end{eqnarray}

 %%%%%%%%%%%%%%%%%%%%%%%%%%%%%%%%%%%%%%%%%%%% Figure 1 %%%%%%%%%%%%%%%%%%%%%%%%%%%%%%%%%%%%%%%%%%%%%%%%%%%%%%%%%%%
\begin{figure}
	\centering 
	\includegraphics[width=12cm, height=8cm, angle=0]{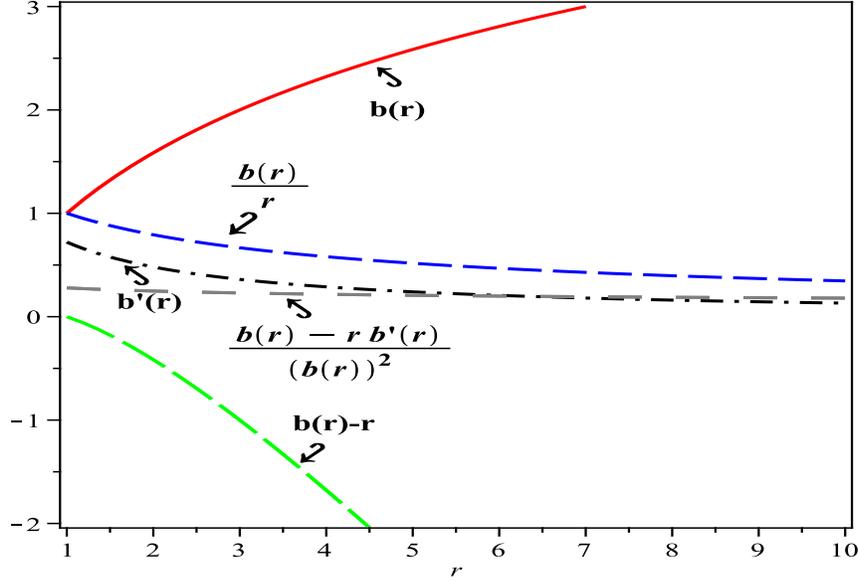}
	\caption {Shape function $b(r)$, flaring-out condition ($b'(r)<1$), throat condition ($b(r)-r<0$), asymptotically flatness condition ($\lim\limits_{r\rightarrow \infty} \frac{b(r)}{r}=0$) against $r$ for throat radius $r_0 = 1.$}
\end{figure} 
  In this article we have discussed different cases based on the values of parameter $\alpha$. Hense there arise three cases for $\alpha$ being positive, negative and zero. We have examined various energy conditions to look into the  geometry of wormholes for these cases. 
 
 The shape function with various conditions  is plotted in Fig.1. As one can see,  all the necessary conditions are satisfied by the given shape function. It can be observed that the geometrical flaring out condition $b'(r)<1$ is validated at the throat of the wormhole. The throat condition $b'(r)<1$ validates the throat radius as $r_0 =1$. Also the asymptotically flatness condition ($\lim\limits_{r\rightarrow \infty} \frac{b(r)}{r}=0$) for throat radius $r_0 = 1$ is verifies the asymptotically flat geometry of universe on both side outside the throat.
 
 %%%%%%%%%%%%%%%%%%%%%%%%%%%%%%%%%%%%%%%%%%%%%%%%%%%
\subsection{Case-I :  $\alpha <0$}
In Figs. 2(a) and 2(b)  we have plotted the energy density $\rho(r)$ and the equation of state parameter $\omega$ against the negative values of $\alpha$ and for throat radius $r_0 = 1$. From Fig. 2(a) the energy density is found to be positive everywhere for $r\geq r_0$ and Fig. 2(b) shows the presence of quintessential matter at the throat i.e. $-1 <\omega<-1/3$ but for away the throat where $r>r_0$ WH EoS is in the phantom region i.e $\omega<-1$. The phantom EoS parameter $\omega <-1$ is well known for Universe will imply in the so-called Big Rip \cite{ref76n1} to avoid such a disastrous scenario have appeared\cite{ref76n2, ref76n3, ref76n4, ref76n5}. Phantom wormholes have also been reported in the literature \cite{ref76n6, ref76n7, ref76n8} albeit it's worth noting that the phantom EoS was summoned rather than obtained from the model in these situations (as in this instance).  From Figs. 3(a) and 3(b) it is clear that both radial and tangential NECs are satisfied for all  $ r\geq r_0$, which implies the validation of WEC. The radial and Tangential DECs are mapped in Figs. 4(a) and 4(b). The radial DEC  is satisfied everywhere but the tangential DEC is violated for $r\geq r_0$. From Fig. 5(a) we can see that strong energy conditions (SEC) are justified everywhere for $ r\geq r_0 = 1$. As for the anisotropy parameter, it is observed from Fig. 5(b) that it is positive everywhere in  $ r\geq r_0$ and for negative values of $\alpha$, which gives the repulsive nature of matter.\\
.

  %%%%%%%%%%%%%%%%%%%%%%%%%%%%%%%%%%%%%%%%%%%%%%%%%%%%%%%%%%%%%%%%%%%%%%%%%%%%% 
    
  %%%%%%%%%%%%%%%%%%%%%%%%%%%%%%%%%%%%%%%%%%%%% Figure 2 %%%%%%%%%%%%%%%%%%%%%%%%%%%%%%%%%%%%%%%%%%%%%%%%%%%%%%%%%%%%
  \begin{figure}
  	(a)\includegraphics[width=8cm, height=8cm, angle=0]{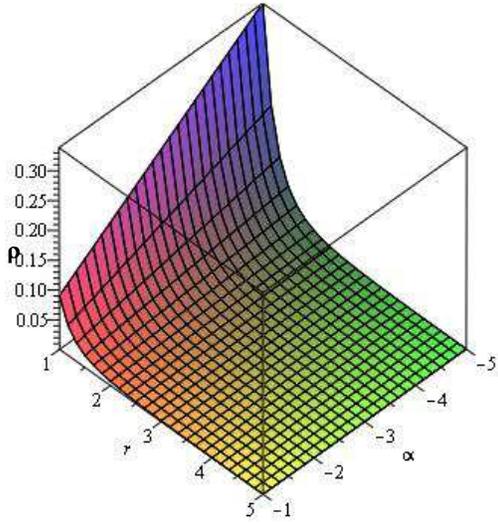}
    (b)\includegraphics[width=8cm, height=8cm, angle=0]{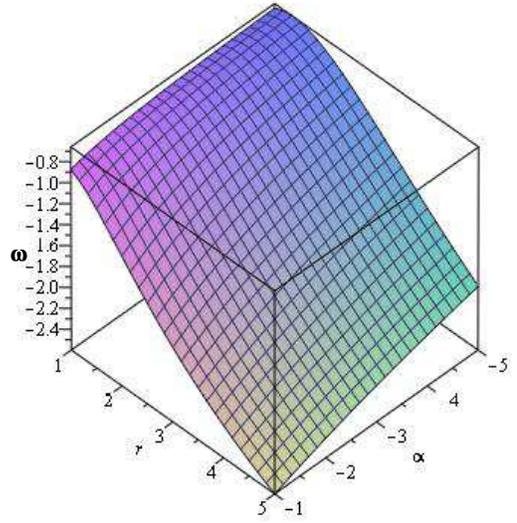}
  	\caption {Variation of Energy density ($\rho(r)$) and  EoS parameter ($\omega$) for throat radius $r_0 = 1$ }
  \end{figure}
  %%%%%%%%%%%%%%%%%%%%%%%%%%%%%%%%% Figure 3 %%%%%%%%%%%%%%%%%%%%%%%%%%%%%%%%%%%%%%%%%%%%%%%%%%%%%%%%%%%%%%%%%%%%%%%%%%%%%%
  
  \begin{figure}
  	(a)\includegraphics[width=8cm, height=8cm, angle=0]{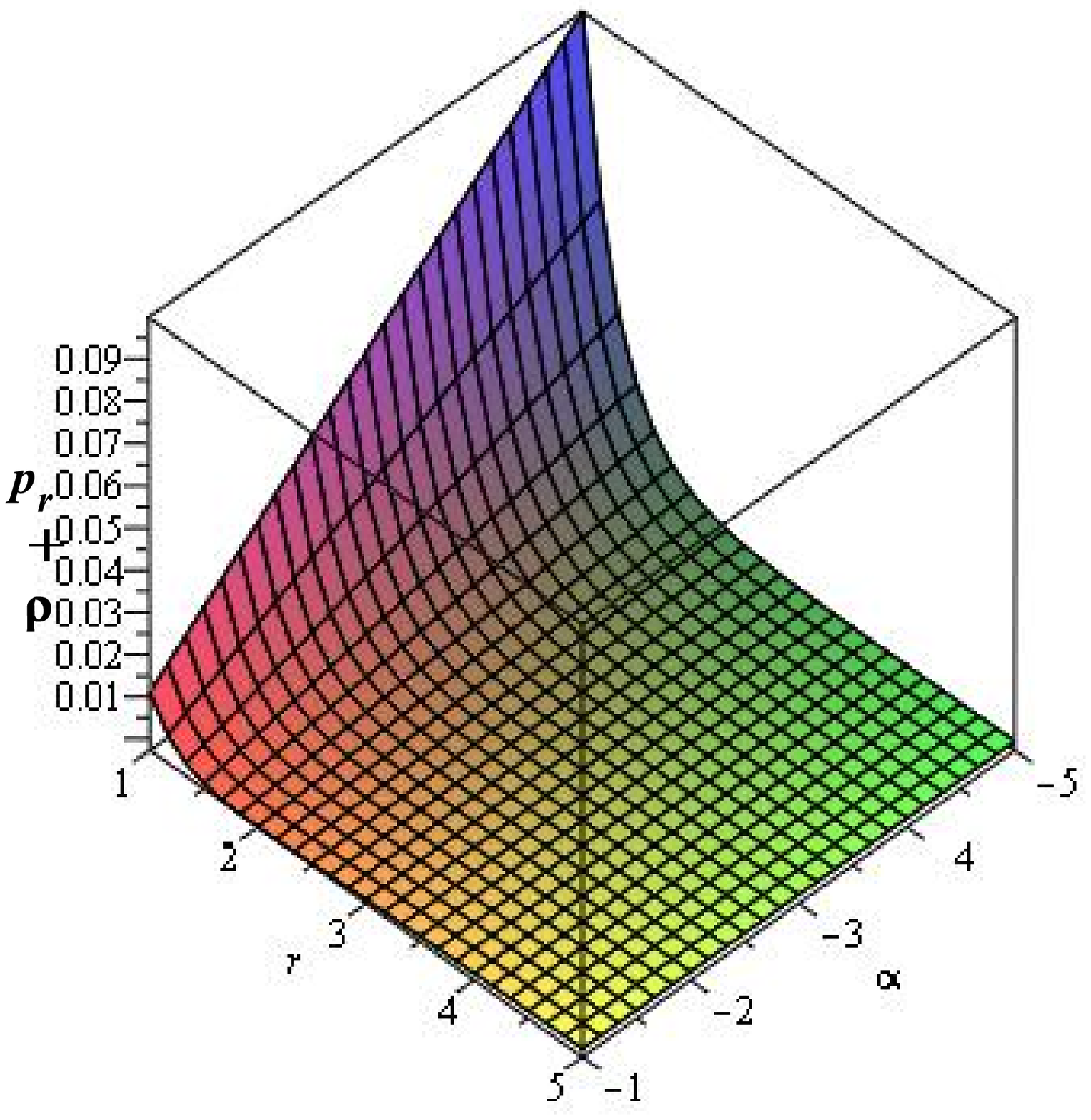}
  	(b)\includegraphics[width=8cm, height=8cm, angle=0]{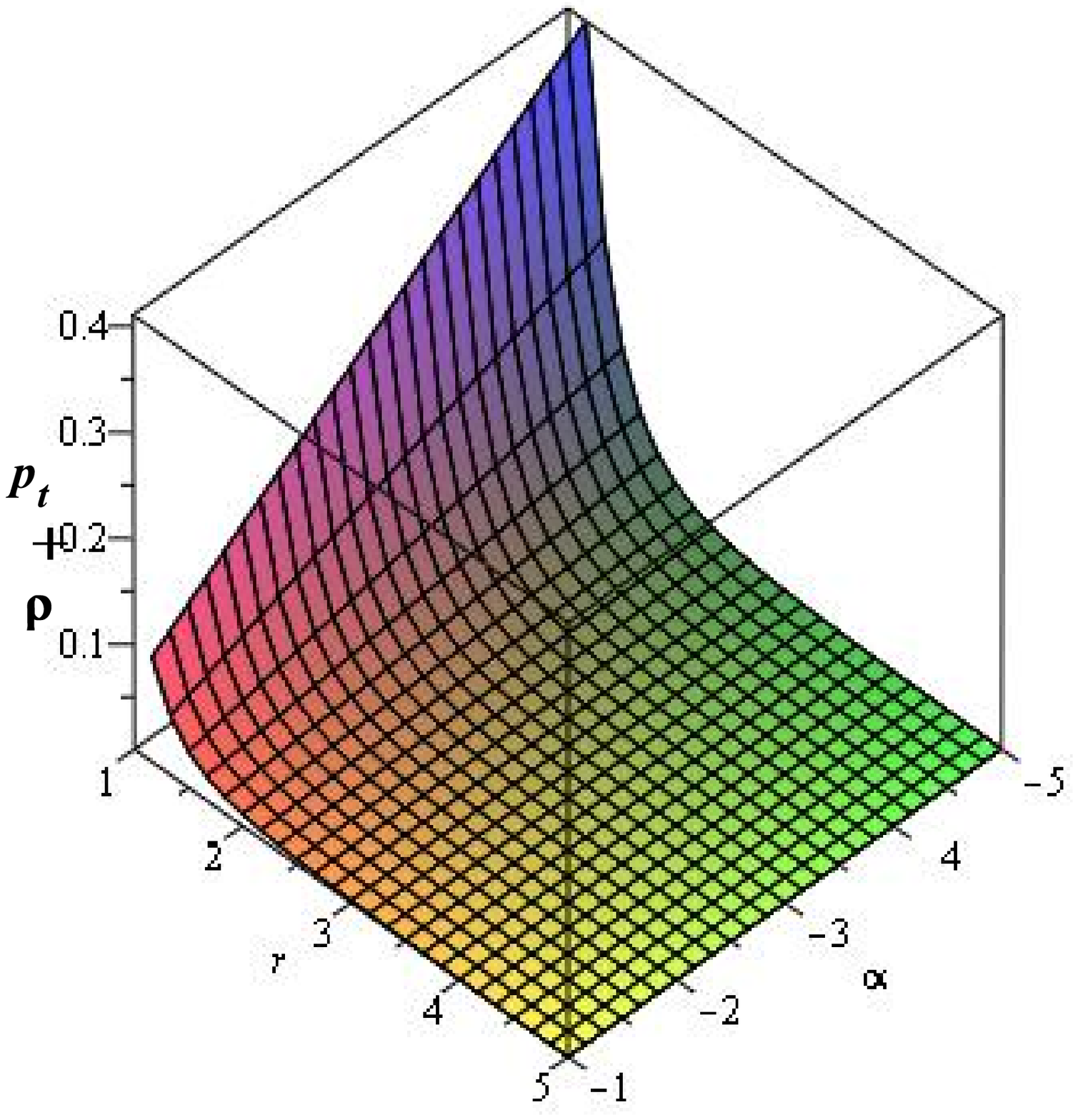}
  	\caption {Variation of NECs ($\rho(r) + p_r(r), \,\rho(r)+p_l(r)$) for throat radius $r_0 = 1$.}
  \end{figure}
  %%%%%%%%%%%%%%%%%%%%%%%%%%%%%%%%%%%%%%%%%%%%%% Figure 4 %%%%%%%%%%%%%%%%%%%%%%%%%%%%%%%%%%%%%%%%%%%%%%%%%%%
  \begin{figure}
  	(a)\includegraphics[width=8cm, height=8cm, angle=0]{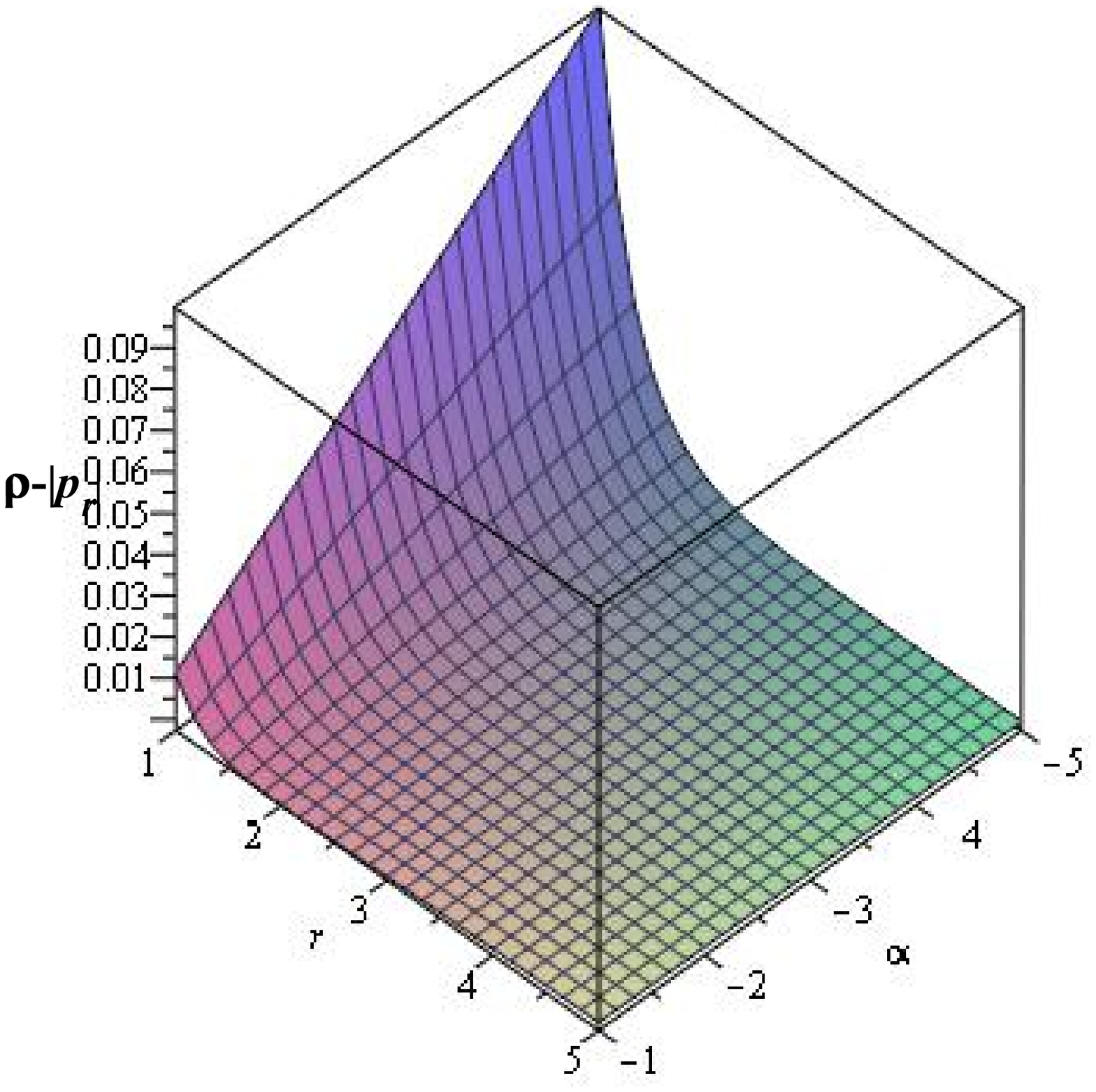}
  	(b)\includegraphics[width=8cm, height=8cm, angle=0]{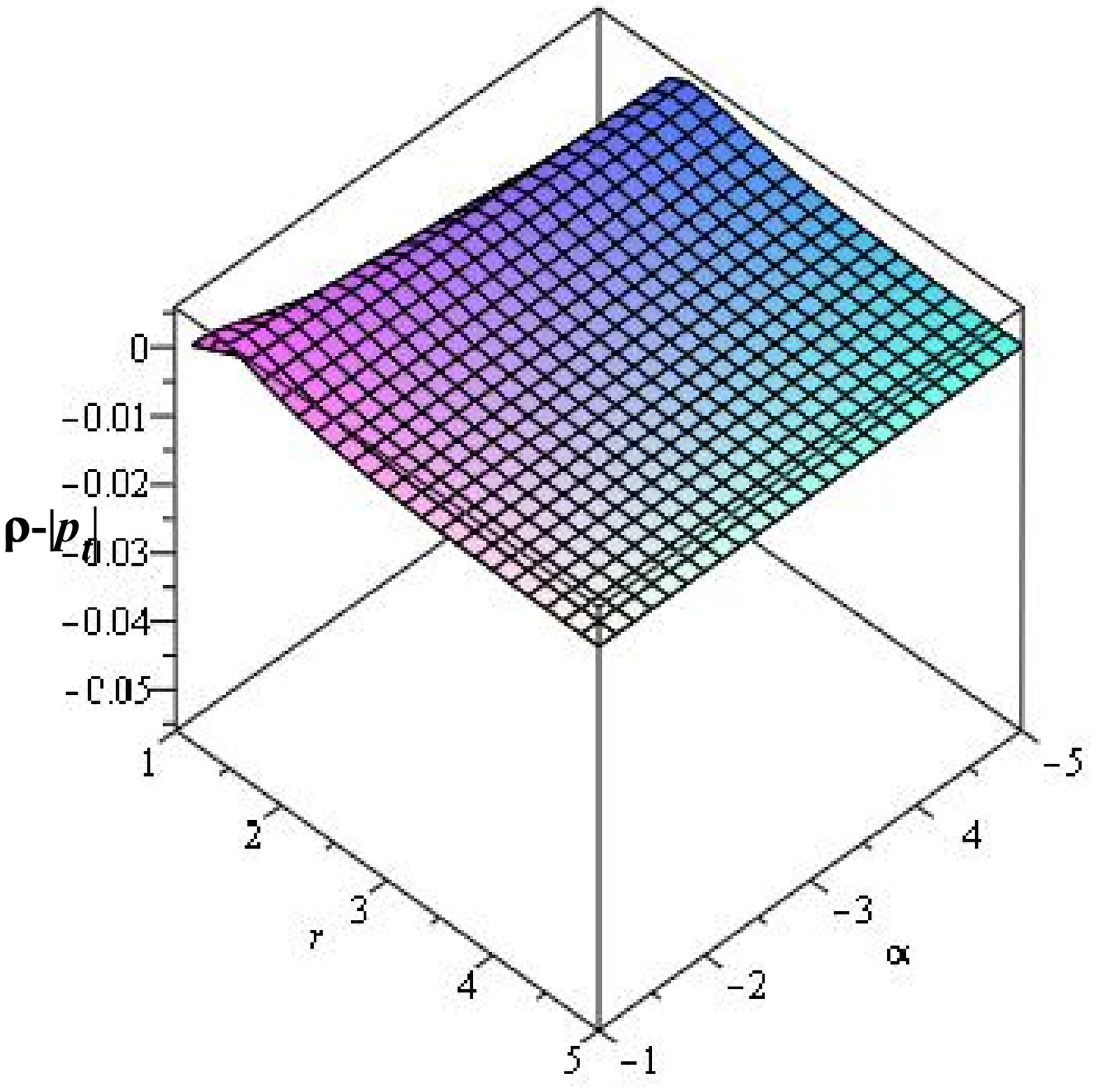}
  	\caption {Variation of DECs ($\rho(r) -|p_r(r)|, \,\rho(r)-|p_l(r)|$) for throat radius $r_0 = 1$.}
  \end{figure}
 
%%%%%%%%%%%%%%%%%%%%%%%%%%%%%%%%%%%%%%%%%%%%%%%%%%%%%%%%%%%% %%%%%%%%%%%%%%%%%%%%Figure5%%%%%%%%%%%%%%% 

  \begin{figure}
  	(a)\includegraphics[width=8cm, height=8cm, angle=0]{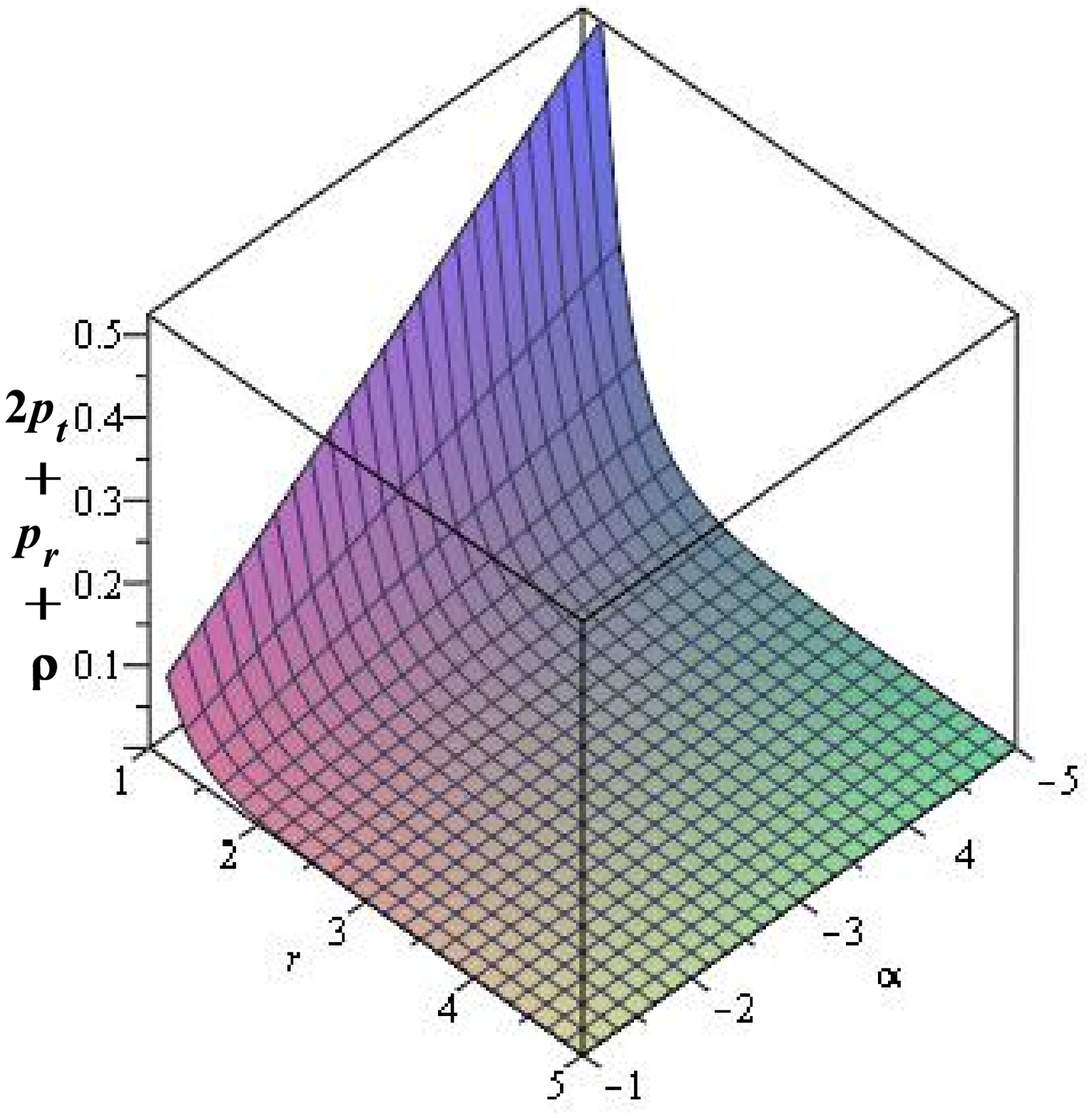}
  	(b)\includegraphics[width=8cm, height=8cm, angle=0]{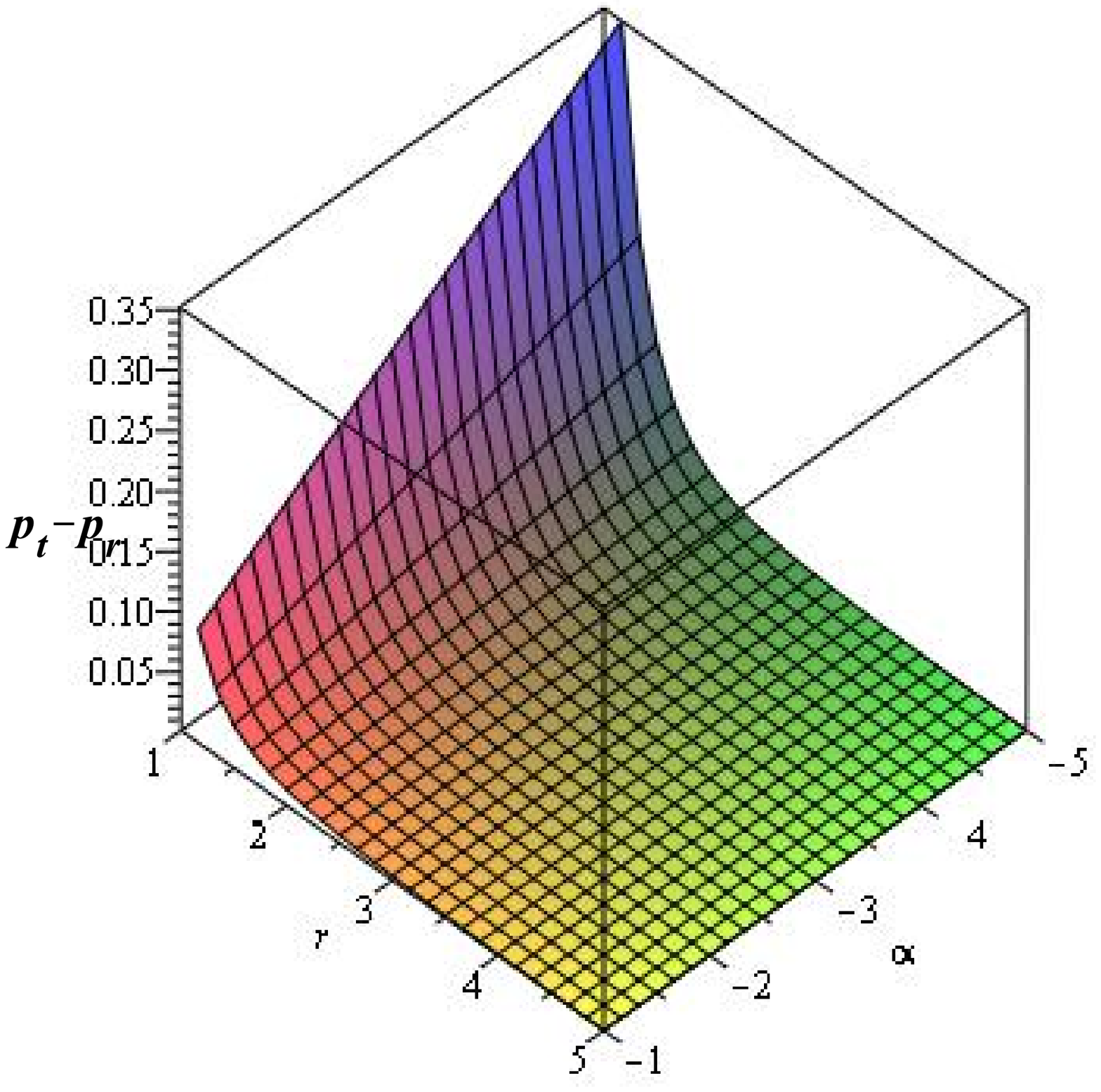}
  	\caption {Variation of SEC ($\rho(r) + p_r(r)+ 2p_l(r)$) and anisotropy parameter ($p_l(r)-p_r(r)$) for throat radius $r_0 = 1$.}
  \end{figure} 
      
     %%%%%%%%%%%%%%%%%%%%%%%%%%%%%%%%%%%%%%%%%%%%%%%%%%%%%%%%%
 %%%%%%%%%%%%%%%%%%%%%%%%%%%%%%%%%%%%%%%%%%%%%%%%%%%%%%%%%%%%  6%%%%%%%%%%%%%%% 
 
 %%%%%%%%%%%%%%%%%%%%%%%%%%%%%%%%%%%%%%%%%%%%%%%%%%%%%%%%%

\subsection{Case-II:  $\alpha> 0$}
 
Here, for positive values of $\alpha $, various energy conditions are analyzed. From Fig. 6(a), the nature of energy density can be analyzed,  which tends to be positive for $r \geq 2$. NECs are both violated in this case as can be seen in plot 7(a) and 7(b). Both the DECs, radial as well as tangential are also not obeyed in the case of positive values of $\alpha$ which are shown in Fig. 8(a) and (b) for every  $r\geq r_0$. The strong energy condition is devised in Fig. 9(a) which indicates the non-validation of SEC for every  $r\geq r_0$. The anisotropy parameter is partially positive but almost negative everywhere else in the given region.  The EoS parameter $\omega$ is plotted in Fig. 6(b) which is negative  everywhere for $r>r_0$ except  near the throat for some particular value of $\alpha$, where it is found to be positive. It also shows the presence of phantom fluid in WH structure. It shows that in general energy conditions are not obeyed by some small and arbitrary quantities at the wormhole throat for $\alpha>0$.

   %%%%%%%%%%%%%%%%%%%%%%%%%%%%Figure6%%%%%%%%%%%%%%%%%%%%%%%%%%%%%%%%
   %%%%%%%%%%%%%%%%%%%%%%%%%%%%%%%%%%%%%%%%%%%%%%%%%%%%%%%%%%%
   
    \begin{figure}
    	(a)\includegraphics[width=8cm, height=8cm, angle=0]{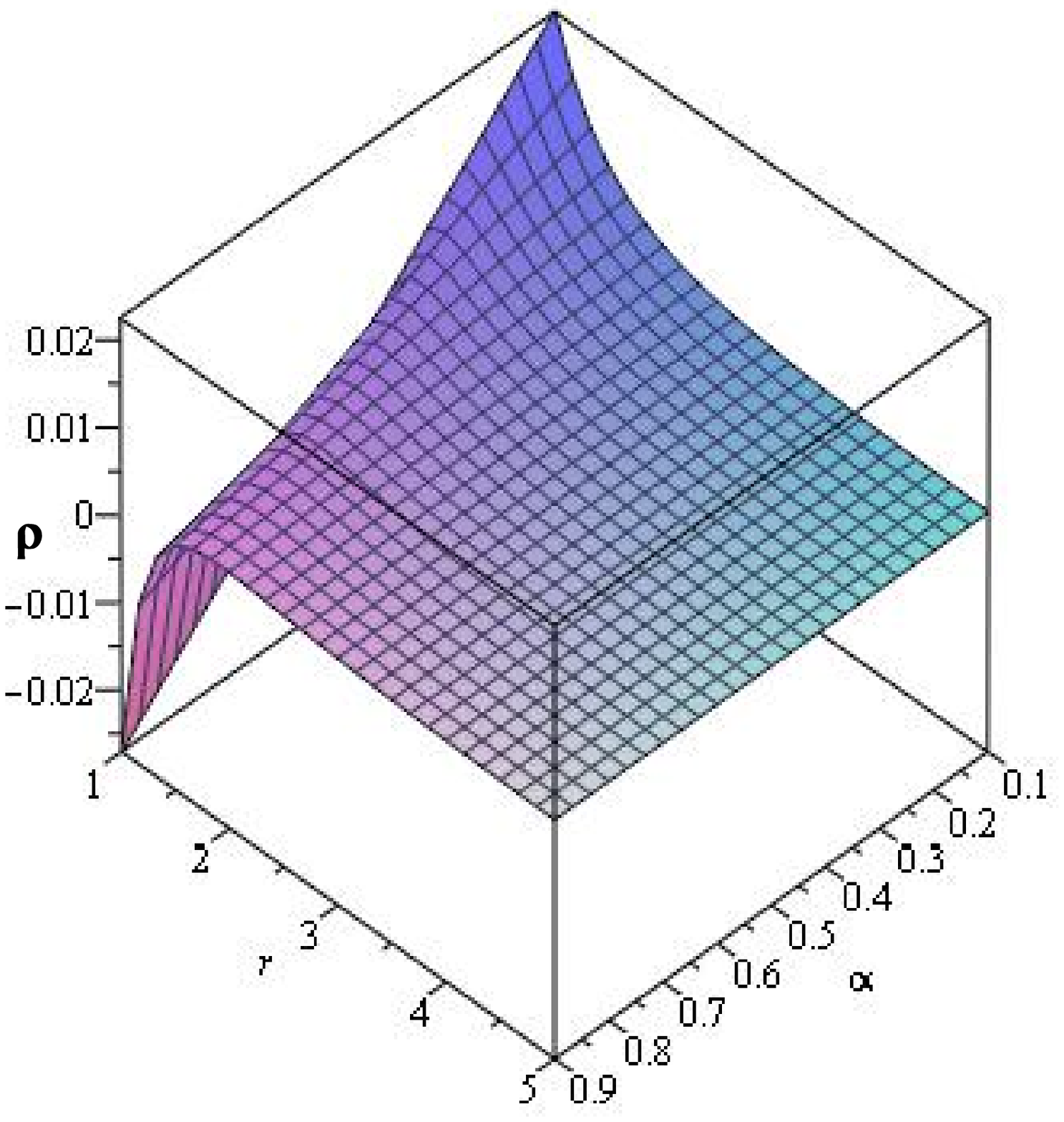}
    	(b)\includegraphics[width=8cm, height=8cm, angle=0]{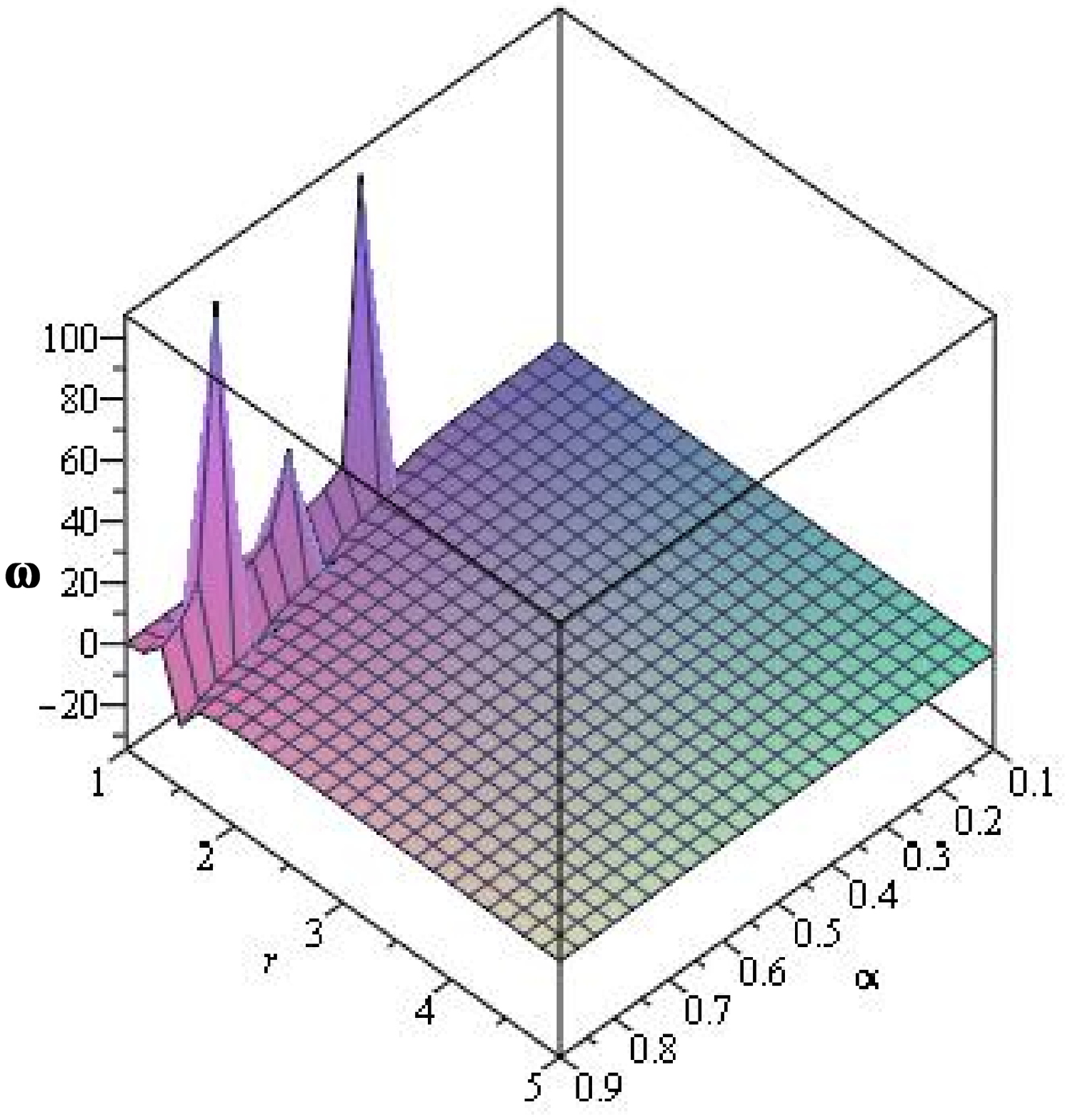}
    	\caption {Variation of Energy density ($\rho(r)$) and  EoS parameter ($\omega$) for throat radius $r_0 = 1$ }
    \end{figure}
    %%%%%%%%%%%%%%%%%%%%%%%%%%%%%%%%% Figure 7 %%%%%%%%%%%%%%%%%%%%%%%%%%%%%%%%%%%%%%%%%%%%%%%%%%%%%%%%%%%%%%%%%%%%%%%%%%%%%%%
    \begin{figure}
    	(a)\includegraphics[width=8cm, height=8cm, angle=0]{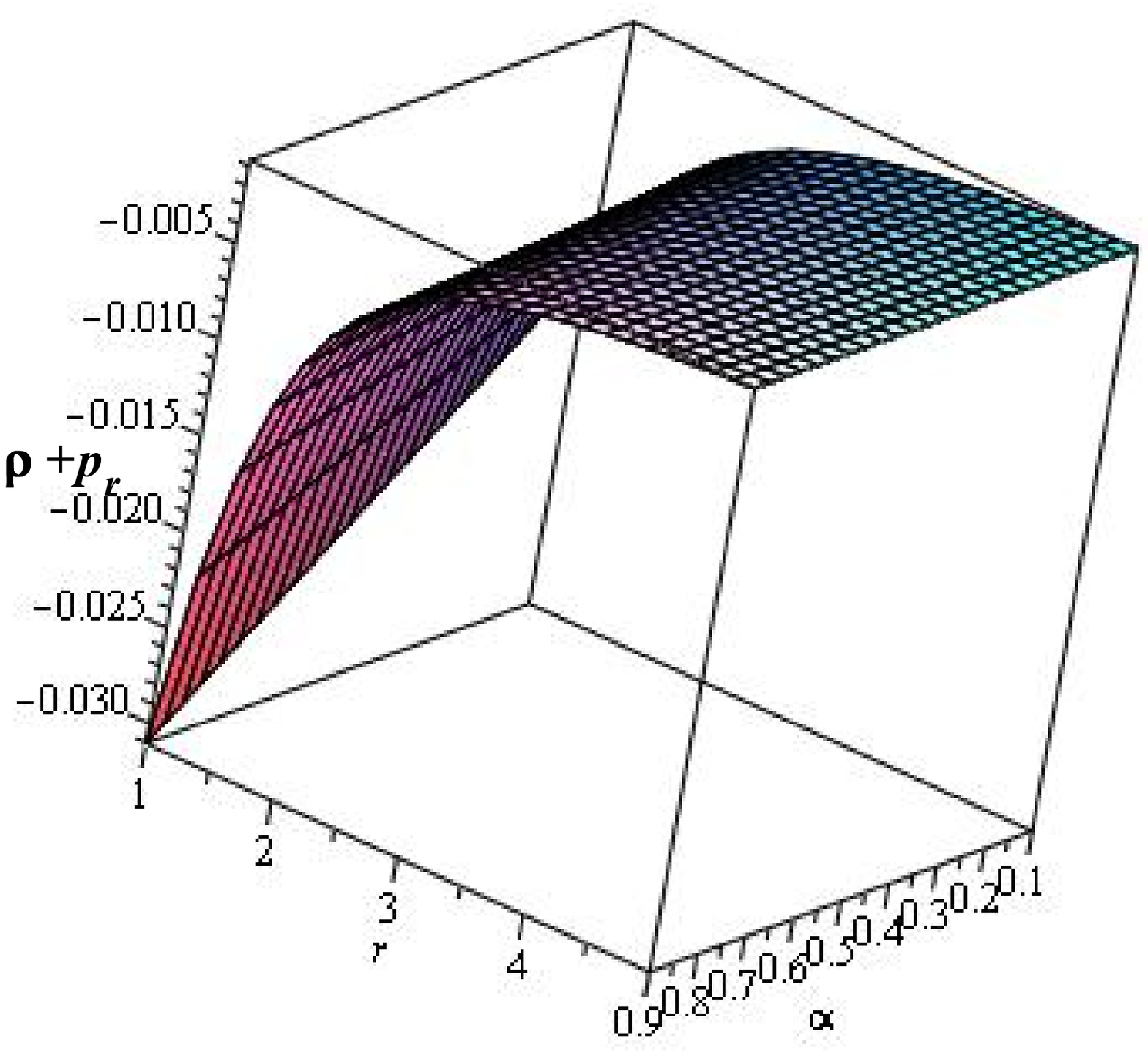}
    	(b)\includegraphics[width=8cm, height=8cm, angle=0]{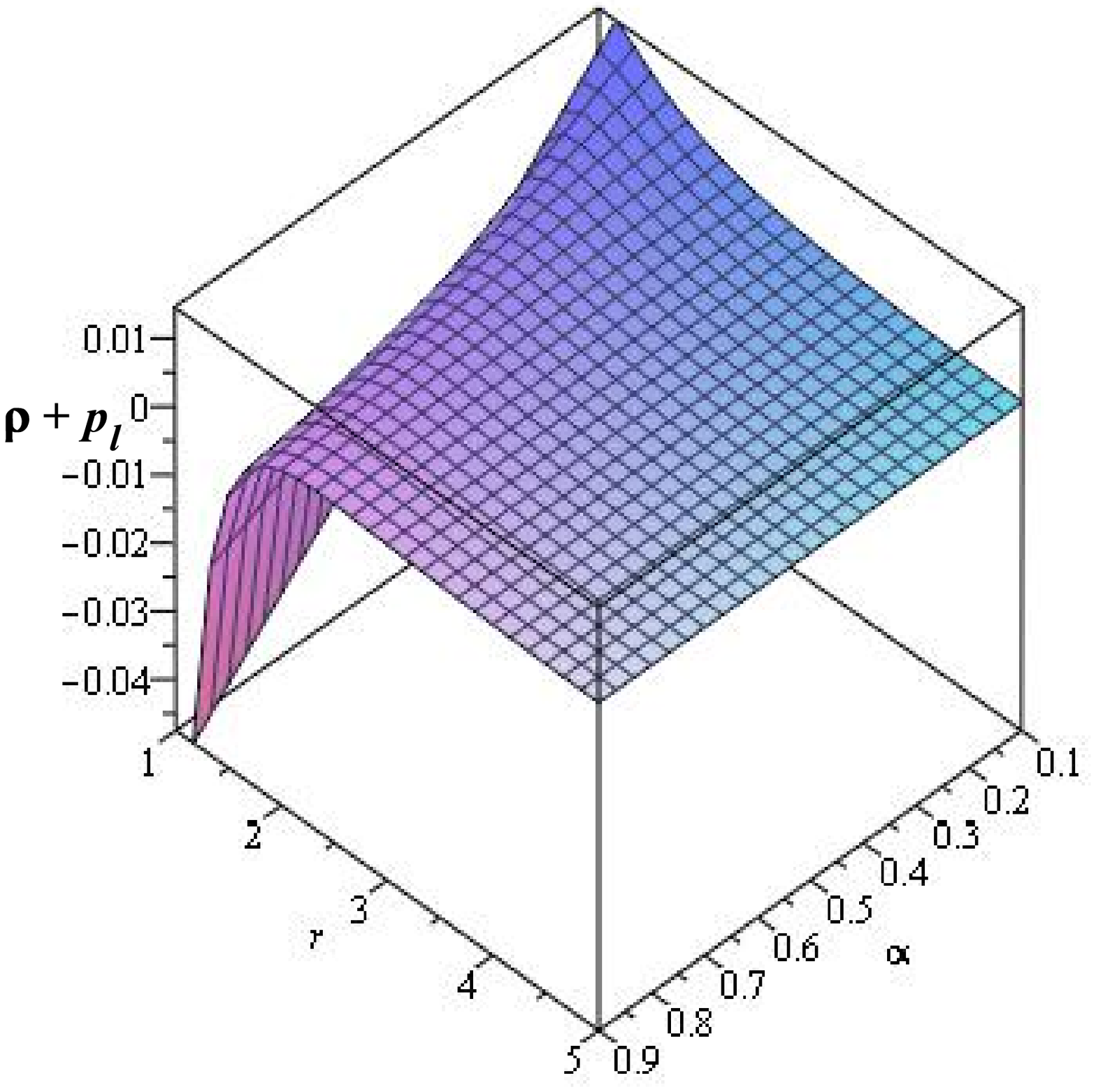}
    	\caption {Variation of NECs ($\rho(r) + p_r(r), \,\rho(r)+p_l(r)$) for throat radius $r_0 = 1$.}
    \end{figure}
    %%%%%%%%%%%%%%%%%%%%%%%%%%%%%%%%%%%%%%%%%%%%%% Figure 8 %%%%%%%%%%%%%%%%%%%%%%%%%%%%%%%%%%%%%%%%%%%%%%%%%%%
    \begin{figure}
    	(a)\includegraphics[width=8cm, height=8cm, angle=0]{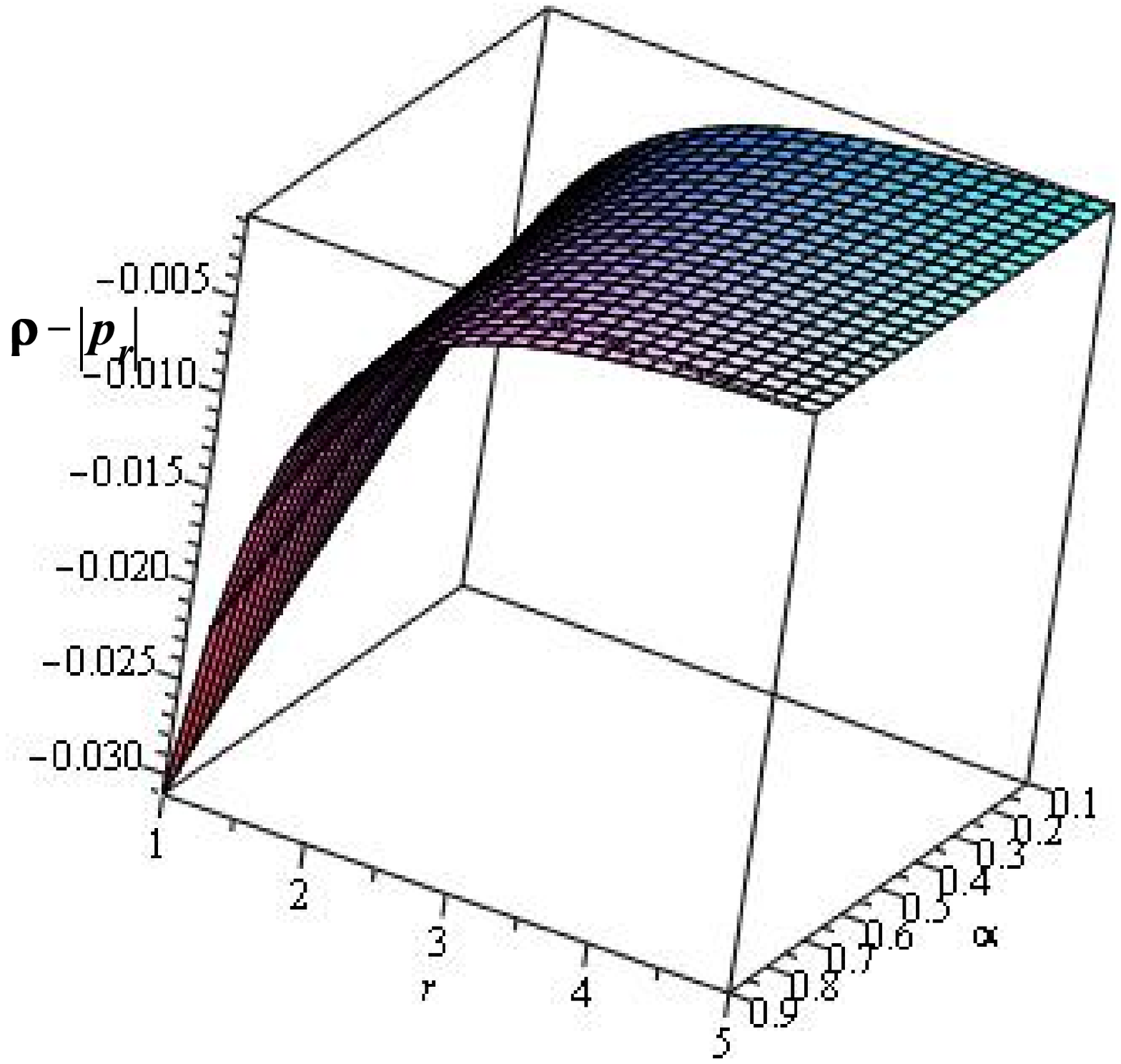}
    	(b)\includegraphics[width=8cm, height=8cm, angle=0]{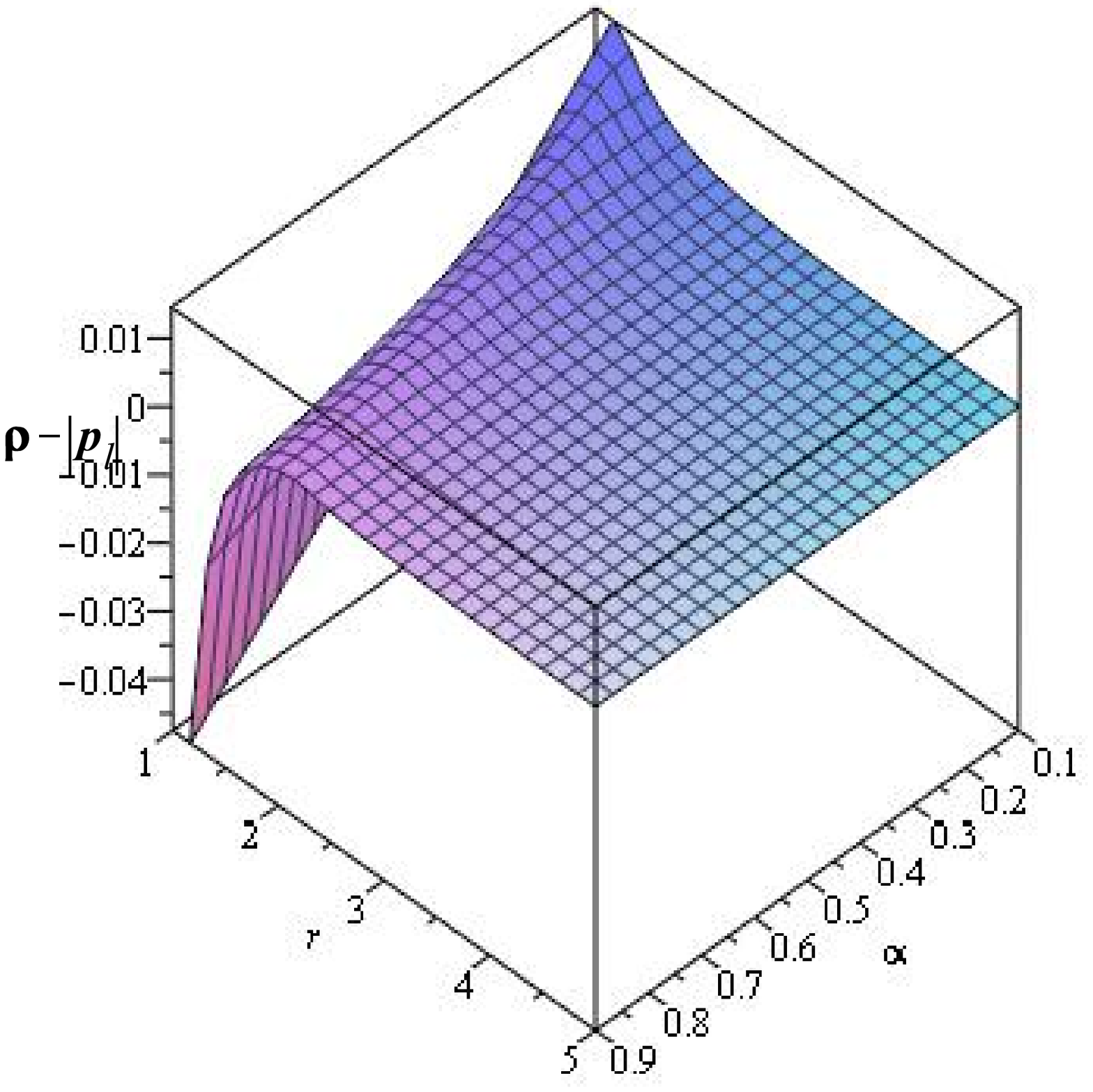}
    	\caption {Variation of DECs ($\rho(r) -|p_r(r)|, \,\rho(r)-|p_l(r)|$) for throat radius $r_0 = 1$.}
    \end{figure}

    %%%%%%%%%%%%%%%%%%%%%%%%%%%%%%%%%%%%%%%%%%%%%%%%%%%%%%%%%%%% %%%%%%%%%%%%%%%%%%%%Figure9%%%%%%%%%%%%%%% 
    \begin{figure}
    	(a)\includegraphics[width=8cm, height=8cm, angle=0]{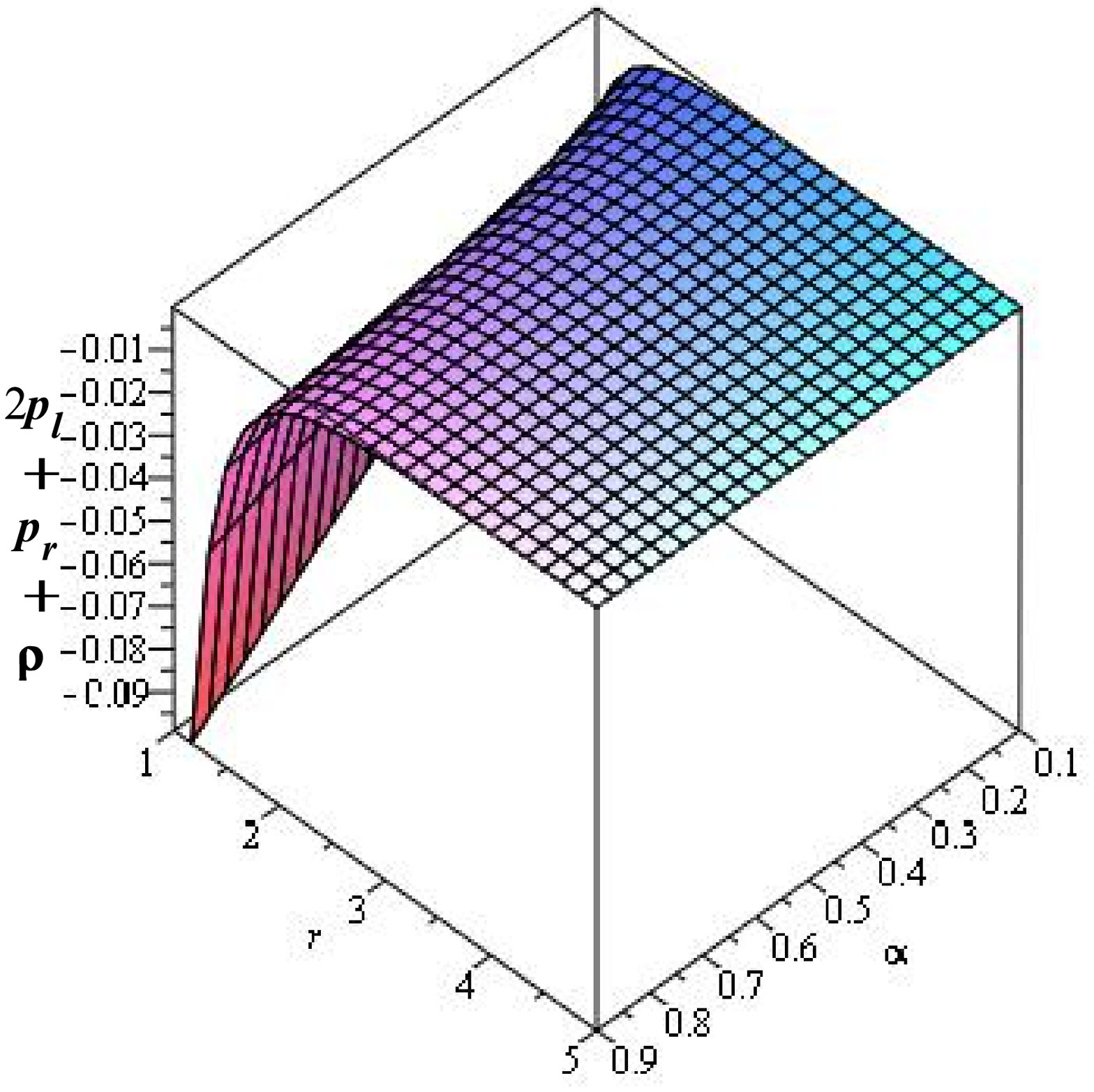}
    	(b)\includegraphics[width=8cm, height=8cm, angle=0]{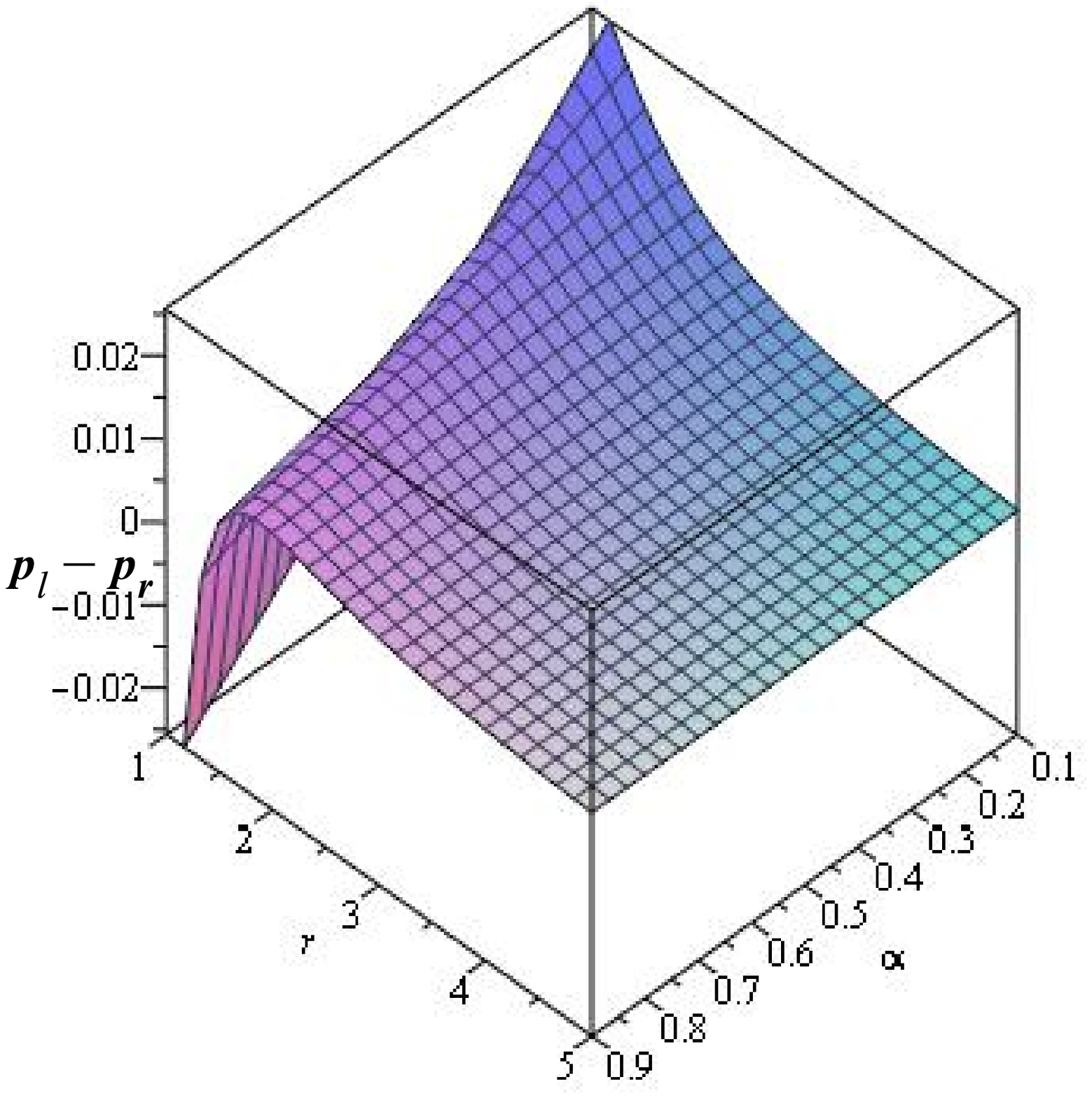}
    	\caption {Variation of SEC ($\rho(r) + p_r(r)+ 2p_l(r)$) and anisotropy parameter ($p_l(r)-p_r(r)$) for throat radius $r_0 = 1$.}
    \end{figure}

%%%%%%%%%%%%%%%%%%%%%%%%%%%%%%%%%%%%%%%%%%%%%%%%%%%%%%%%%%%%%%%%%%%%

\subsection{Case-III : $\alpha = 0$ i.e. general relativity} 

The energy conditions for $\alpha = 0$ are judged by plotting them in Figures (12- 13). Energy density $\rho(r)$, here comes out to be positive  for all $r>r_0$. Figs. 11(a) and 11(b) show that the NEC in terms of radial pressure is not authenticated while that in terms of tangential pressure is validated everywhere for all $r>r_0$. The radial DEC is violated in Fig. 12(a) while the tangential DEC is satisfied everywhere in the region. The validity of SEC is also not established as shown in Fig. 13(a). The anisotropy parameter  is positive for all $r>r_0$ for $\alpha = 0 $, which indicates the repulsive nature of WH geometry. Fig. 10(b) shows the values of the EoS parameter which is $< -1$ as $r \geq r_0$ which again indicate the presence of phantom fluid.  
Thus all the energy conditions are violated and WH geometry is repulsive which reduces exactly to Morris-Thorne of general relativity. \\

%%%%%%%%%%%%%%%%%%%%%%%%%%%%%%%%%%%%%%%%%%%%%%%%%%%%%%%%%%%%%%%%%%%%%%%%%%%%% 

%%%%%%%%%%%%%%%%%%%%%%%%%%%%%%%%%%%%%%%%%%%%% Figure 10 %%%%%%%%%%%%%%%%%%%%%%%%%%%%%%%%%%%%%%%%%%%%%%%%%%%%%%%%%%%%
\begin{figure}
	(a)\includegraphics[width=8cm, height=8cm, angle=0]{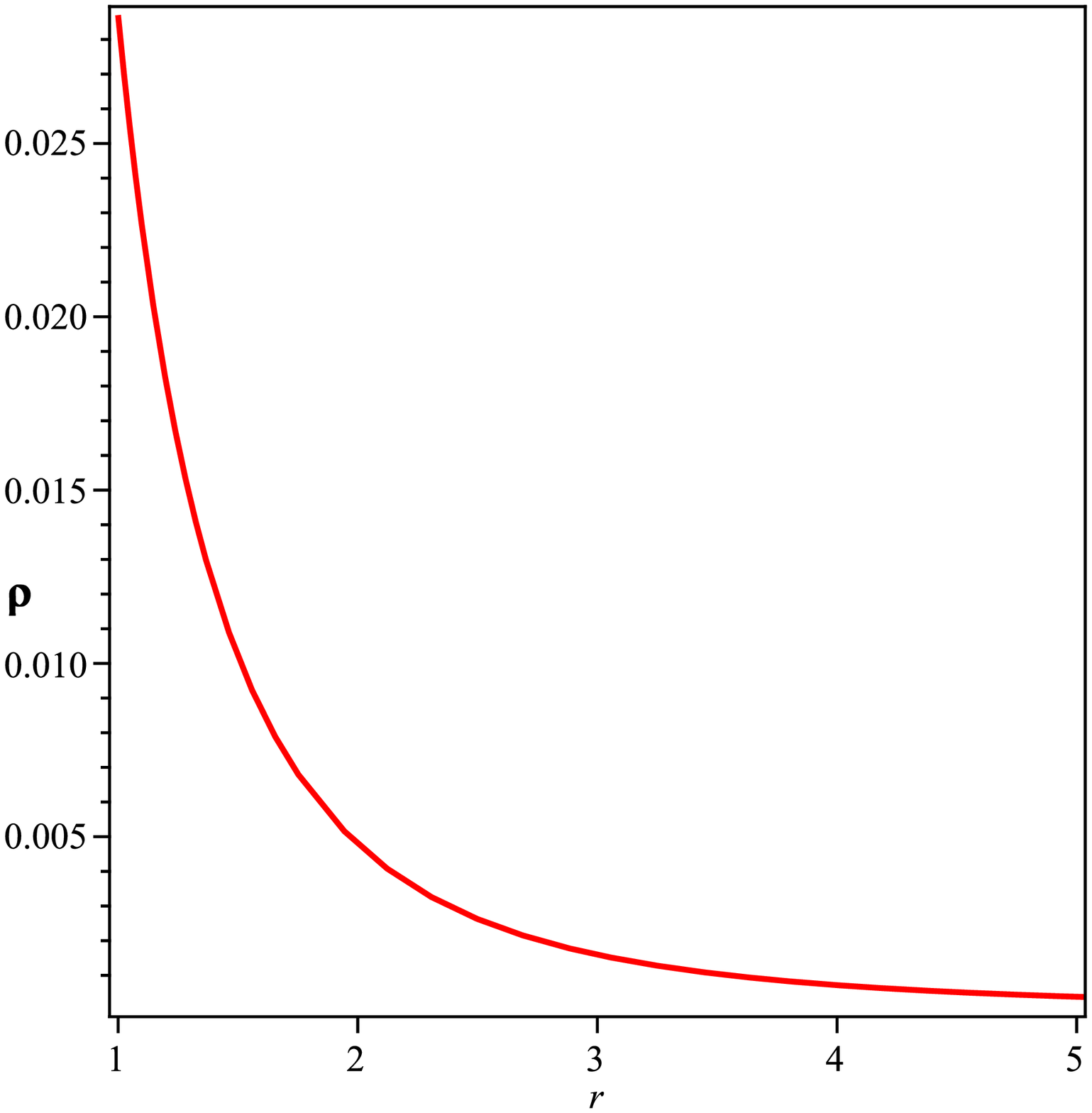}
	(b)\includegraphics[width=8cm, height=8cm, angle=0]{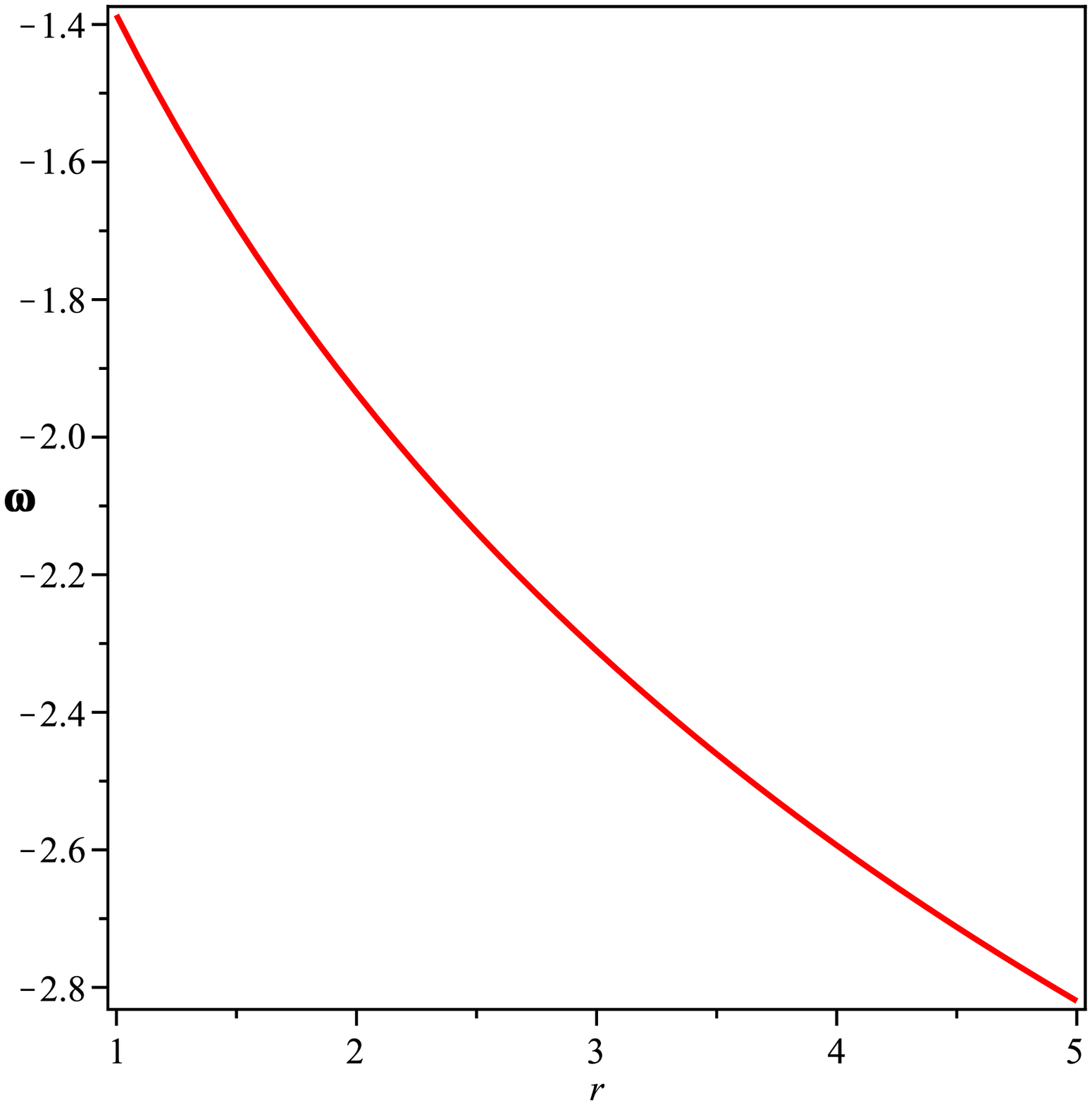}
	\caption {Variation of Energy density ($\rho(r)$) and  EoS parameter ($\omega$) for throat radius $r_0 = 1$ }
\end{figure}
%%%%%%%%%%%%%%%%%%%%%%%%%%%%%%%%% Figure 11 %%%%%%%%%%%%%%%%%%%%%%%%%%%%%%%%%%%%%%%%%%%%%%%%%%%%%%%%%%%%%%%%%%%%%%%%%%%%%%%
\begin{figure}
	(a)\includegraphics[width=8cm, height=8cm, angle=0]{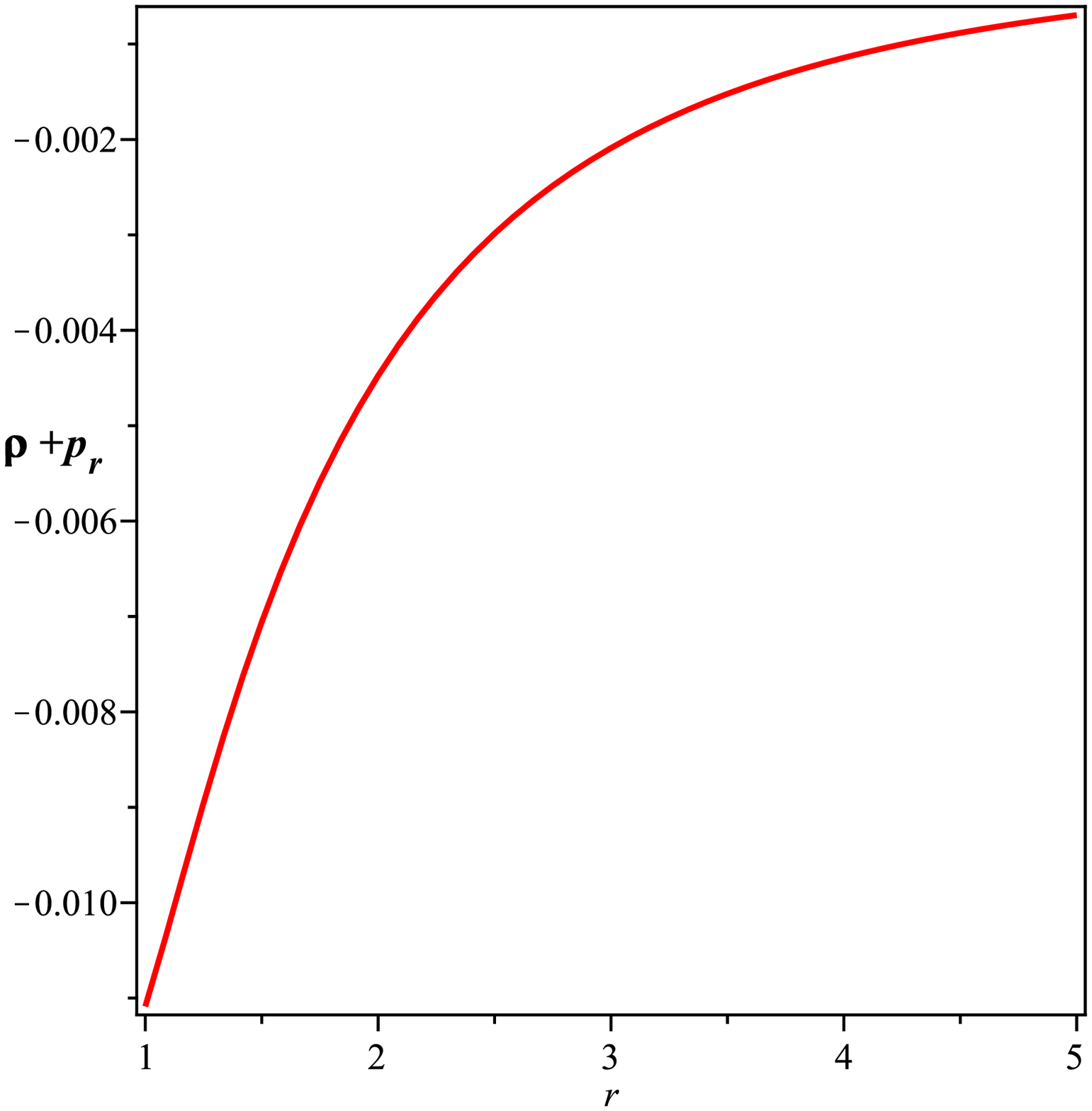}
	(b)\includegraphics[width=8cm, height=8cm, angle=0]{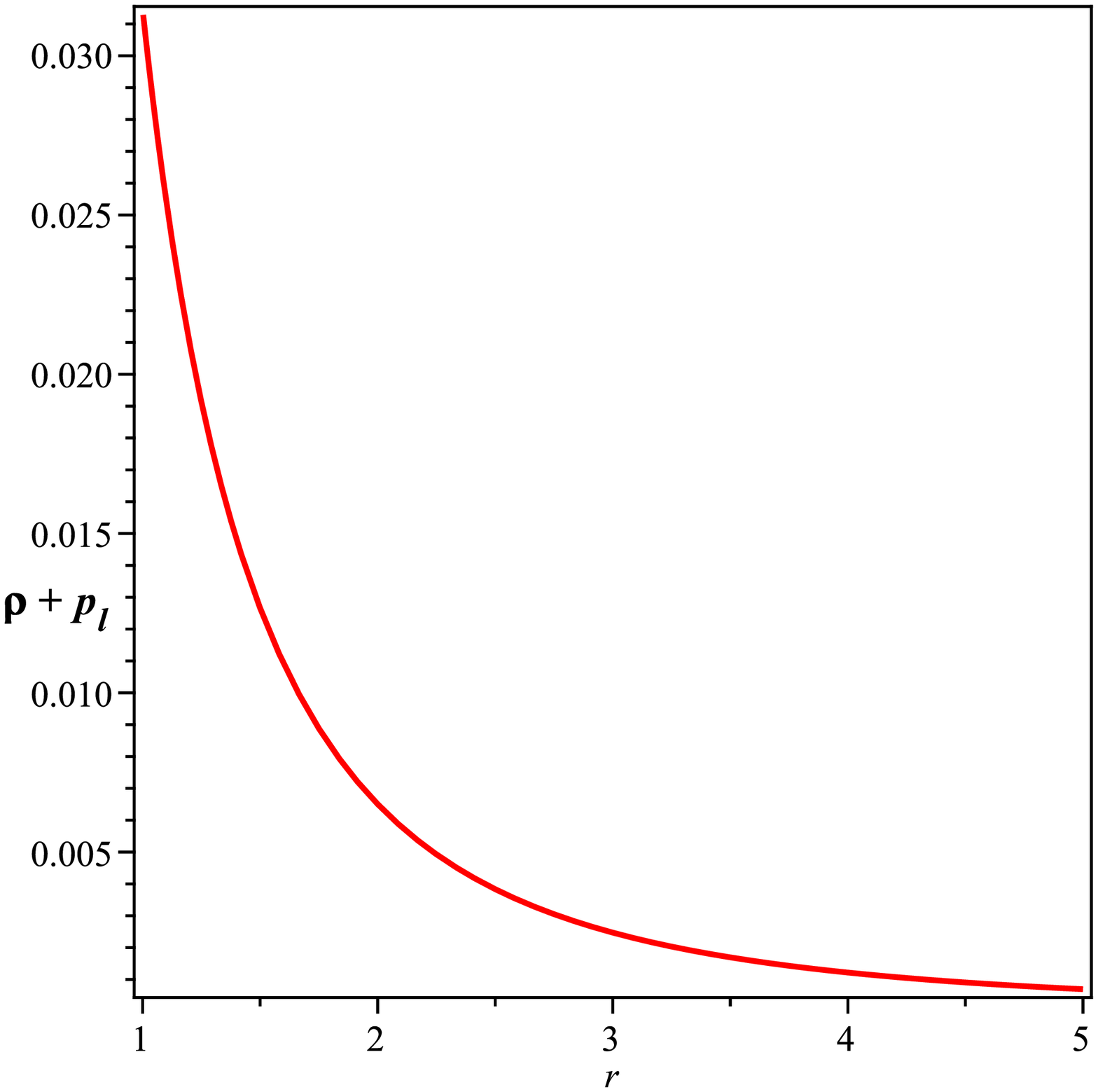}
	\caption {Variation of NECs ($\rho(r) + p_r(r), \,\rho(r)+p_l(r)$) for throat radius $r_0 = 1$.}
\end{figure}
%%%%%%%%%%%%%%%%%%%%%%%%%%%%%%%%%%%%%%%%%%%%%% Figure 12 %%%%%%%%%%%%%%%%%%%%%%%%%%%%%%%%%%%%%%%%%%%%%%%%%%%
\begin{figure}
	(a)\includegraphics[width=8cm, height=8cm, angle=0]{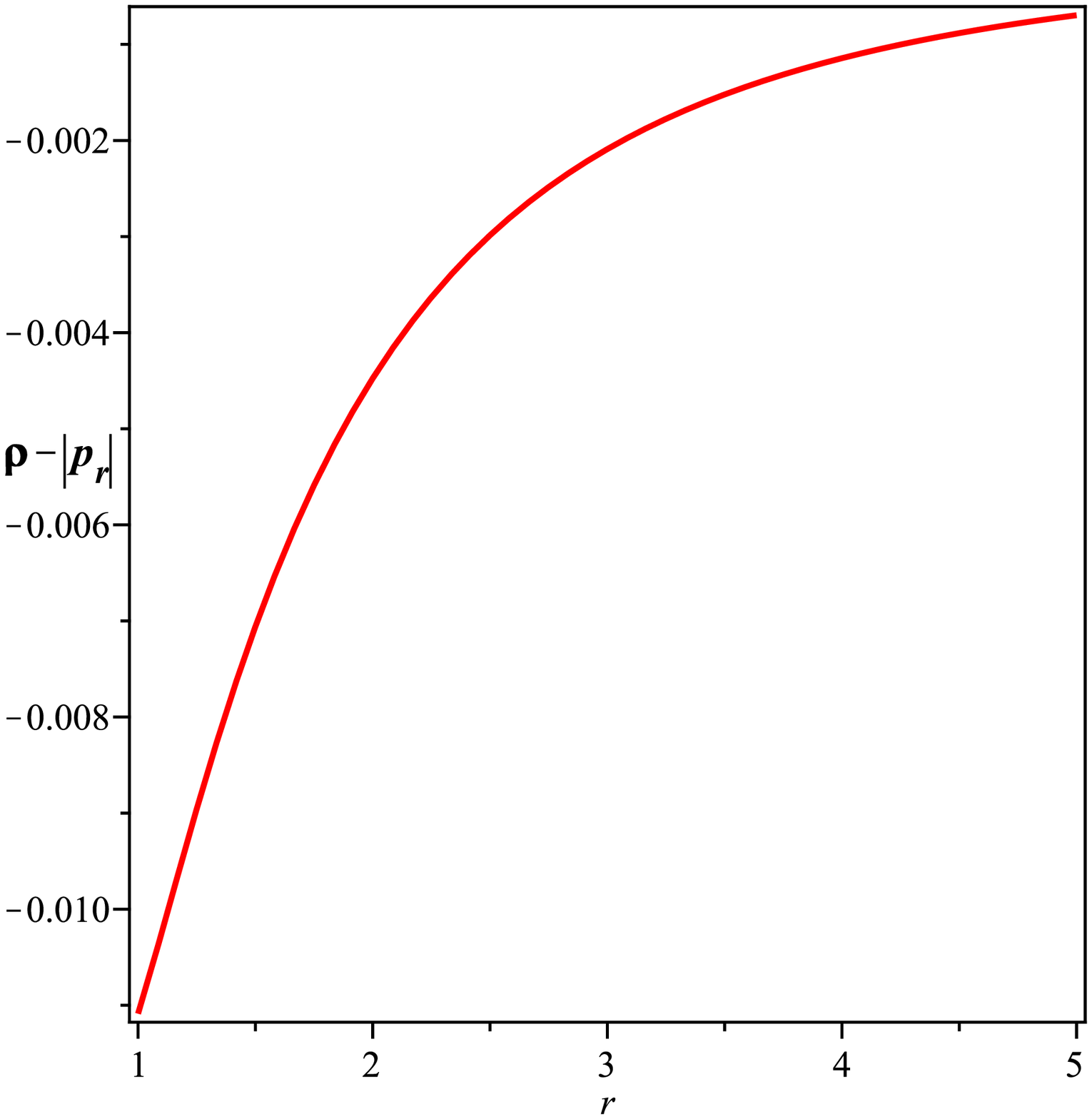}
	(b)\includegraphics[width=8cm, height=8cm, angle=0]{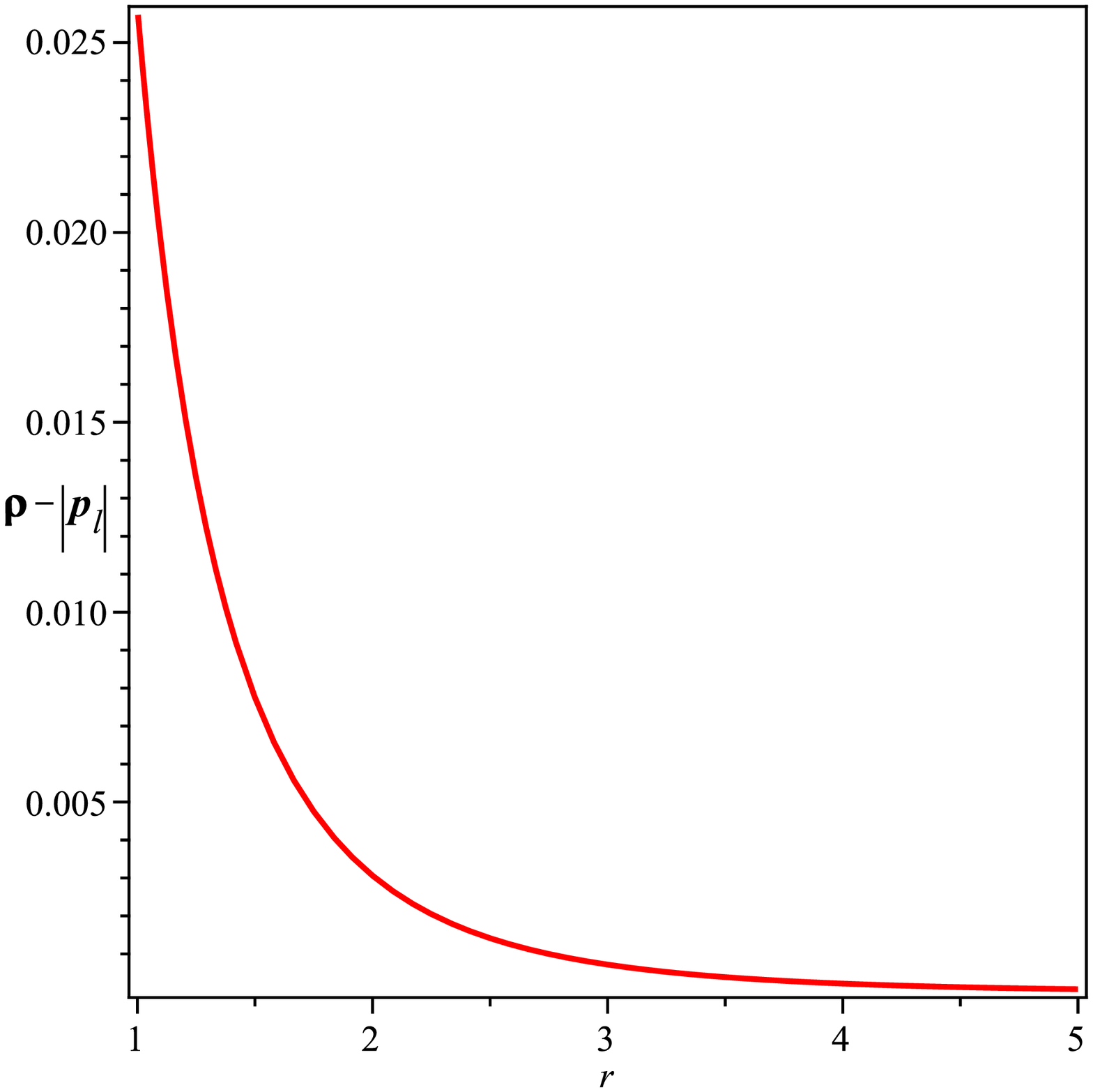}
	\caption {Variation of DECs ($\rho(r) -|p_r(r)|, \,\rho(r)-|p_l(r)|$) for throat radius $r_0 = 1$.}
\end{figure}

%%%%%%%%%%%%%%%%%%%%%%%%%%%%%%%%%%%%%%%%%%% Figure13
 %%%%%%%%%%%%%%%%%%%%  
\begin{figure}
	(a)\includegraphics[width=8cm, height=8cm, angle=0]{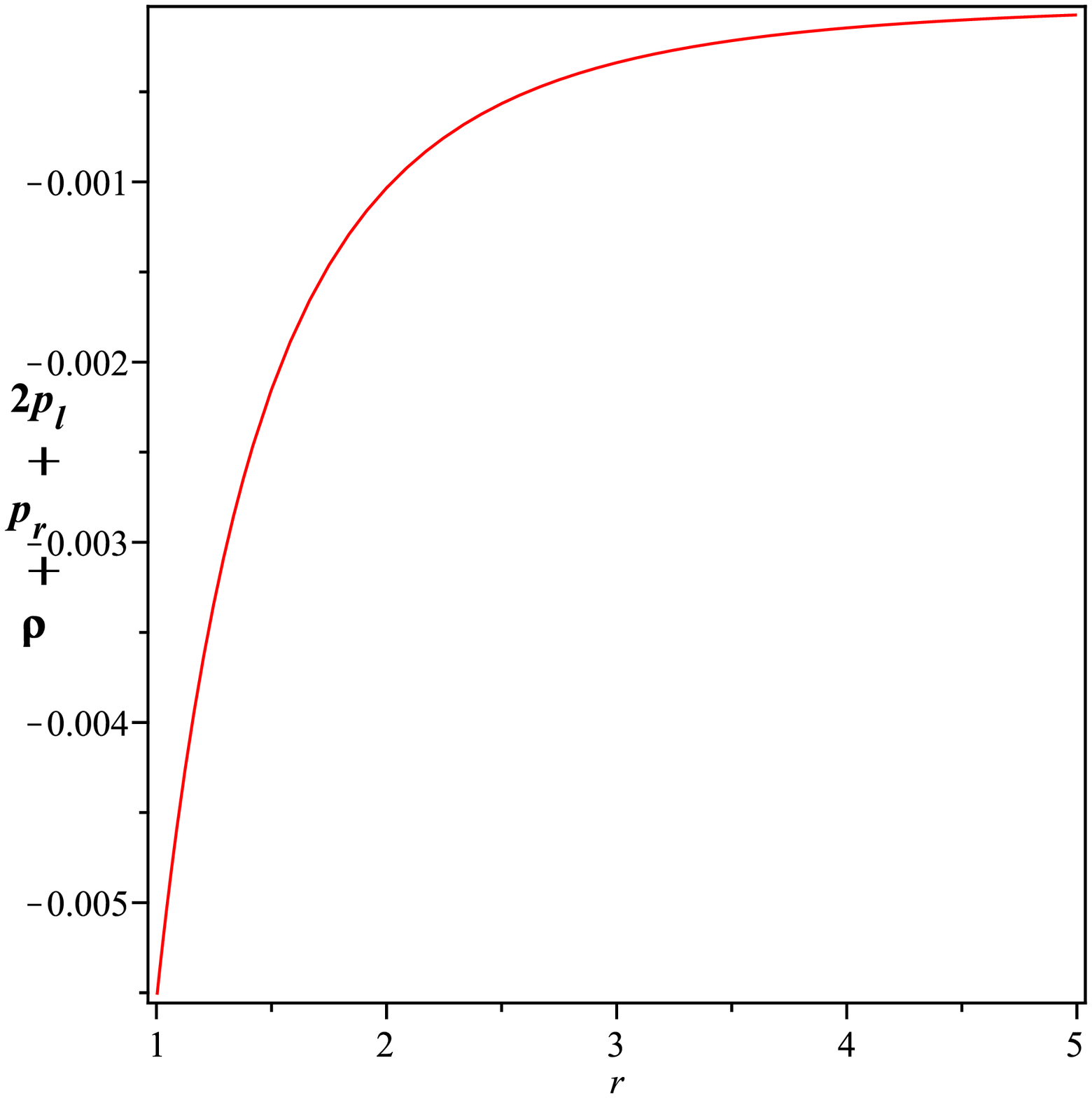}
	(b)\includegraphics[width=8cm, height=8cm, angle=0]{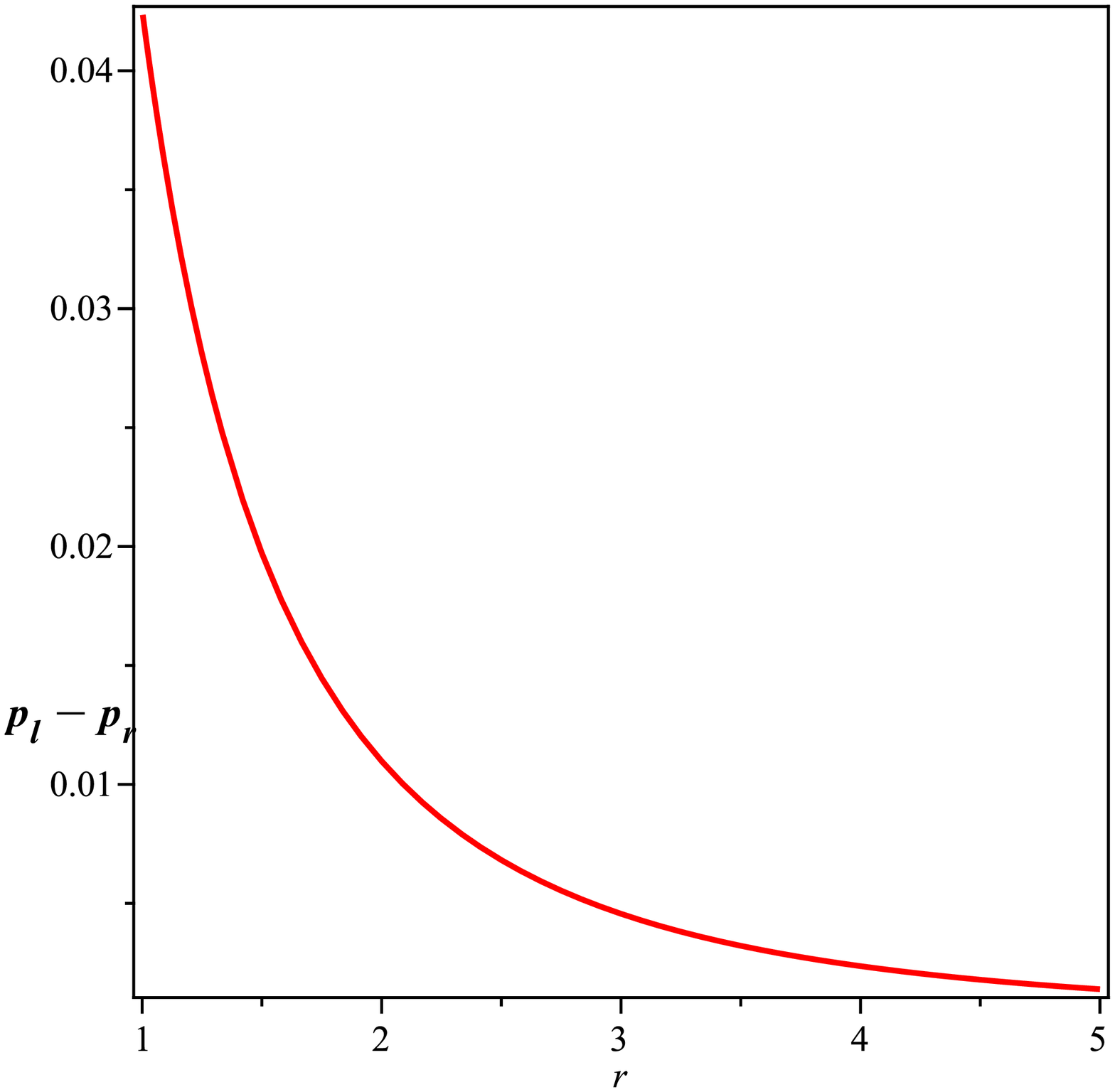}
	\caption {Variation of SEC ($\rho(r) + p_r(r)+ 2p_l(r)$) and anisotropy parameter ($p_l(r)-p_r(r)$) for throat radius $r_0 = 1$.}
\end{figure} 

%%%%%%%%%%%%%%%%%%%%%%%%%%%%%%%%%%%%%%%%%%%%%%%%%%%%%%%%%
%%%%%%%%%%%%%%%%%%%%%%%%%%%%%%%%%%%%%%%%%%%%%%%%%%%%%%%%%%%%  6%%%%%%%%%%%%%%%

%%%%%%%%%%%%%%%%%%%%%%%%%%%%%%%%%%%%%%%%%%%%% Figure 14 %%%%%%%%%%%%%%%%%%%%%%%%%%%%%%%%%%%%%%%%%%%%%%%%%%%%%%%%%%%%
\begin{figure}
	(a)\includegraphics[width=8cm, height=8cm, angle=0]{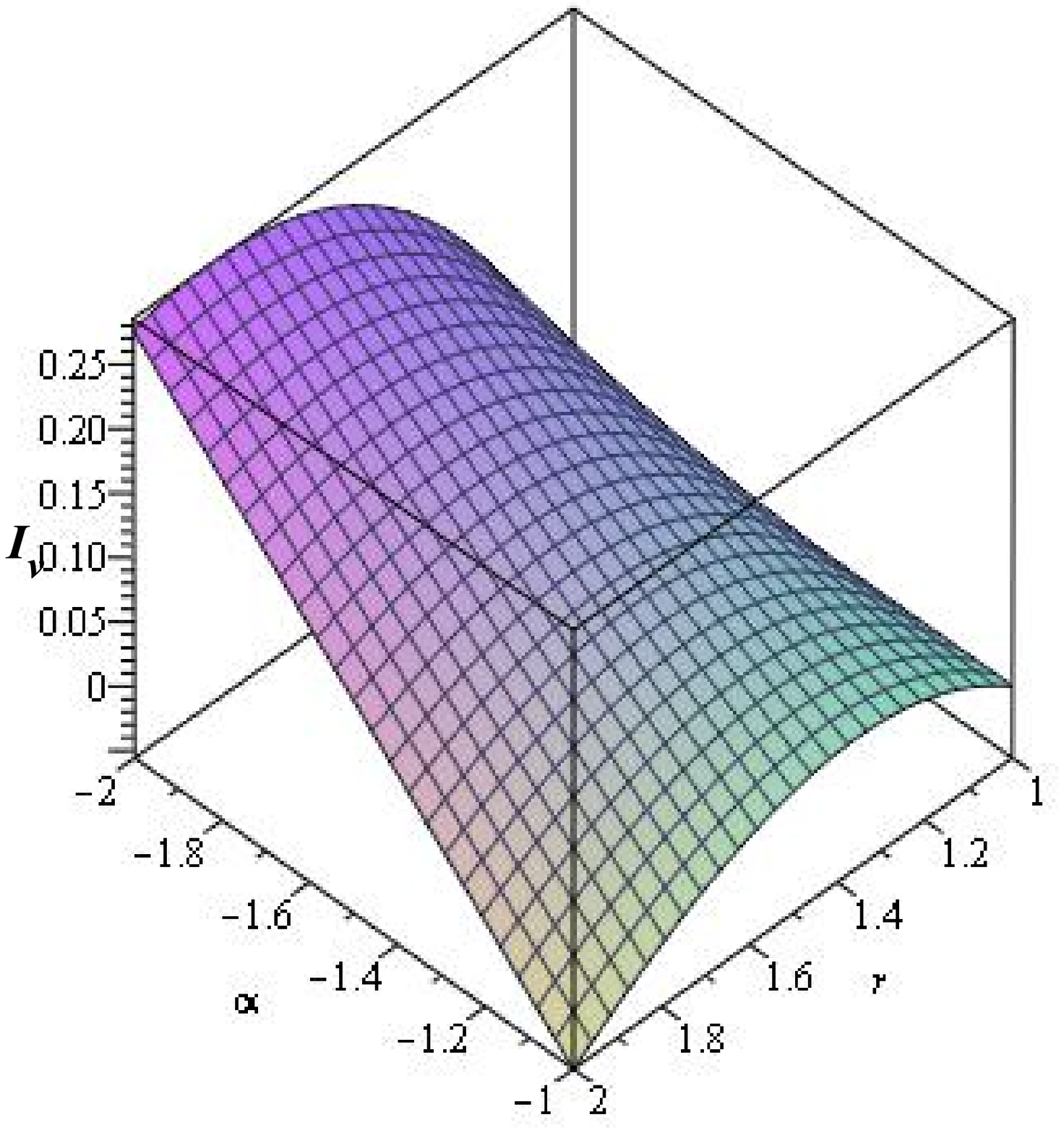}
	(b)\includegraphics[width=8cm, height=8cm, angle=0]{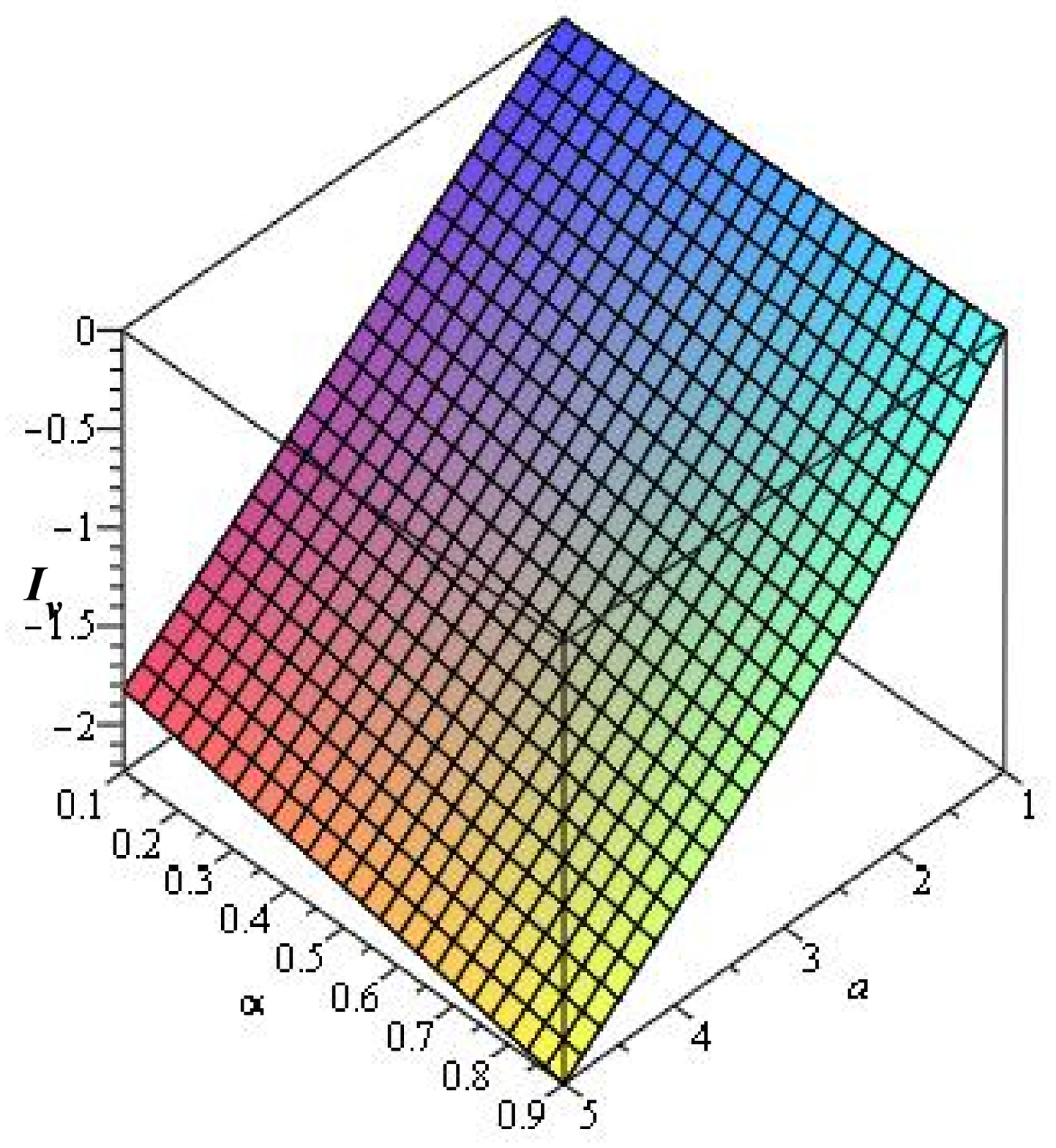}
	\begin{center}
		(c)\includegraphics[width=8cm, height=8cm, angle=0]{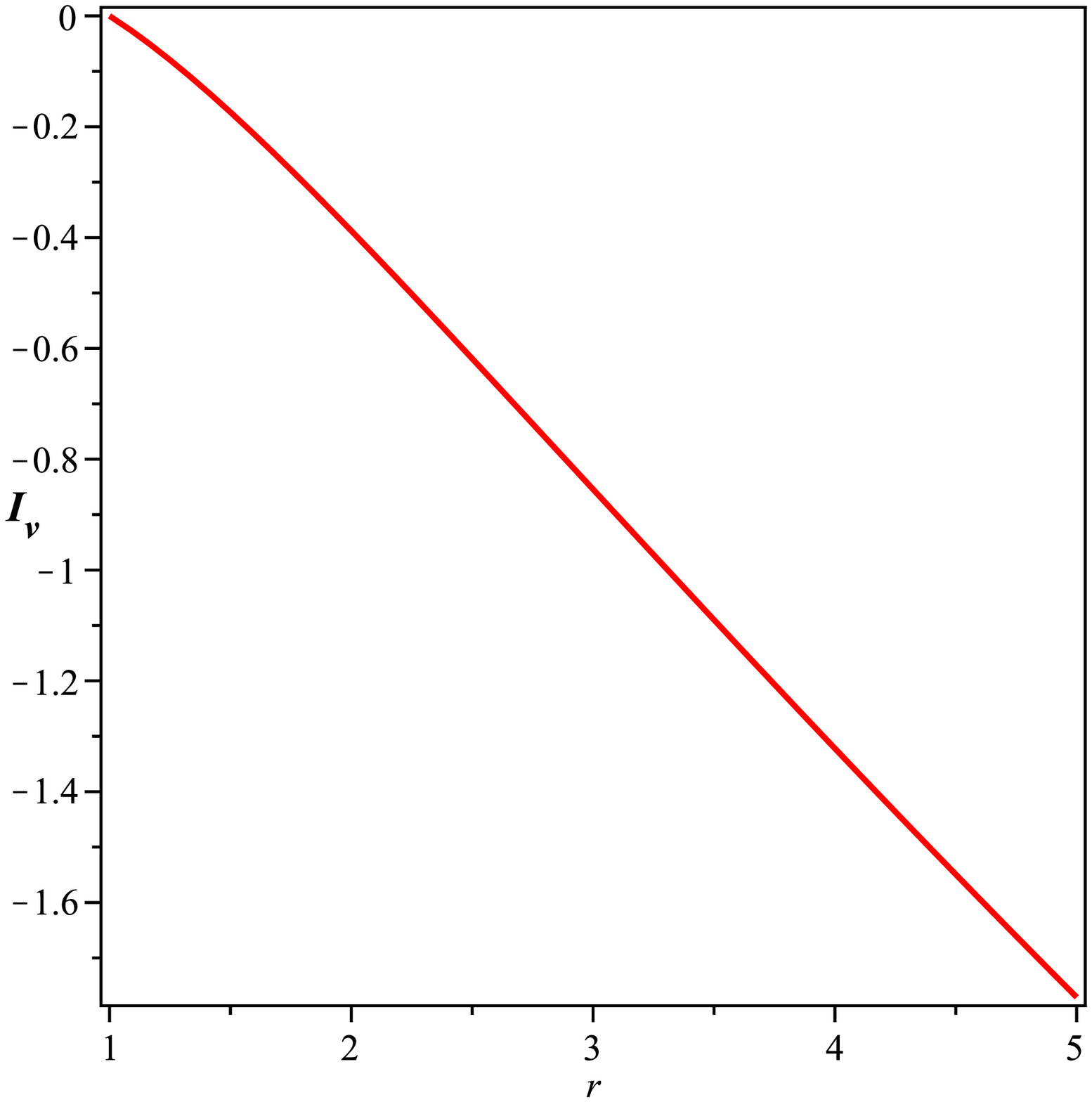}
	\end{center}
	\caption {Volume integral quantifier against (a) $\alpha < 0 $ , (b) $\alpha < 0$  and (c) $\alpha = 0$ where $r_0 = 1$ }
\end{figure}
%%%%%%%%%%%%%%%%%%%%%%%%%%%%%%%%
%%%%%%%%%%%%%%%%%%%%%%%%%%%%%%%%%%%%%%%%%% 

\section{Volume Integral Quantifier}

The infringement of averaged null energy conditions is required to hold up the traversable wormholes but only a small quantity of exotic matter required to hold the wormhole open, can be achieved by considering a pertinent wormhole geometry. This knowledge about the quantity of non-standard matter present, is perceived by the volume integral quantifier or VIQ for which the total integrals $\int T_{\eta\zeta}k^{\eta}k^{\zeta}$   and $\int T_{\eta\zeta}U^{\eta}U^{\zeta}$ have to be solved.Here $U^{\eta}$ is the four velocity. \cite{ref16,ref77}.
The integral used to calculate the VIQ is the definite integral given by

\begin{equation}\label{eq19}
I_{v} = 2 \int_{r_0}^\infty {\left[\rho\left(r\right) + p_{r}{\left(r\right)}\right]} dV  
\end{equation}
where, $ V = r^{2}Sin\theta dr d\theta d\phi$. Further the amount of non-standard matter can also be measured by rewriting this volume integral as 
   
   \begin{equation}\label{eq20}
   	I_{v} = 8 \pi \int_{r_0}^\infty {\left[\rho\left(r\right) + p_{r}{\left(r\right)}\right]} r^{2} dr	
\end{equation}

In Figs. 14(a), 14(b) and 14(c), we have plotted the values of VIQ for different cases based on values of $\alpha$. From Fig. 14(a), it can be observed that for negative values of $\alpha$, the value of integral tends to zero when $r\rightarrow +r_0$ and  is positive and increases initially but  as the values of $r$ increase it comes out to be negative. It indicates the presence of exotic matter in a small amount in the throat and presence of non-exotic matter in a good amount away from the throat. In Fig 14(b), VIQ is mapped against positive values of $\alpha$. It is observed from the figure that the value of integral is negative and is also tending to zero as $r\rightarrow r_0$. Which also indicates that only a small amount of exotic matter can be obtained within the throat by considering suitable wormhole geometry. Fig 14(c) gives a glimpse of the nature of VIQ in case of $\alpha = 0$ i.e. GR. It is clear from the graph that in case of GR the VIQ i.e. $ I_v \rightarrow -\infty$, which again shows that by a peculiar choice of wormhole geometry the quantity of exotic matter can be limited to the amount that is required to keep the wormhole open.

%%%%%%%%%%%%%%%%%%%%%%%%%%%%%%%%%%%%%%%%%%%%%%%%%%%%%%%%%%%%%%%%%%%%%%%%%%%%%%%
\section{Conclusion}
The general limitation in GR to have wormhole geometry is the violation of averaged energy conditions in a vast region of WH or in entire geometry. In the present paper we have examined the static and spherically symmetric traversable WHs for 4D-EGB gravity. Throughout our discussion we have taken shape function $b(r)={\frac{r_{0}\ln(r+1)}{\ln({r_0}+1)}}$ and radial dependent redshift function  $\phi(r)=\ln  \left( {\frac {r_{{0}}}{r}}+1 \right)$ to explore the possible WH solutions in 4D-EGB gravity against different values of $\alpha$. From case-I we can summarize that for negative values of $\alpha$ both the NEC, WEC are not violated. SEC is also justified for this case. Radial DEC is satisfied while tangential DEC is violated. The anisotropy parameter is positive everywhere showing the non-attractive geometry. VIQ, for this case, is tending to zero $r\rightarrow +r_o$ i.e at the throat from which we can conclude that a small amount of exotic matter can be present in the wormhole even if NECs are not violated, contradictory to GR. When we increase the value of $r$ we get non-exotic matter in the wormhole geometry. The value of the equation of state parameter $\omega$ also indicates that there is the presence of quintessential matter at the throat and phantom fluid outside the throat in the WH.

Case-II examines the validity of energy conditions for positive values of $\alpha$. Since both the null energy conditions are violated which in turn leads to the violation of weak energy conditions. This violation of NEC can be interpreted as the existence of exotic matter in the throat but for some values $r$ energy density is positive and tangential NEC is also satisfied which suggests that there is non-exotic matter near the throat radius. SEC is also violated throughout in case-II which identifies with the inflation of the Universe hence authenticating the presence of exotic matter in the WH. Radial DEC is violated throughout while tangential DEC is satisfied in small regions of WH.

As depicted in case-III, the WH geometry for $\alpha =0$  in EGB behaves like the one in GR. The radial NEC is violated and tangential NEC is satisfied for all $r>r_0$. Energy density is positive everywhere, which implies that WEC is violated, indicating the existence of exotic matter. SEC is violated everywhere and the anisotropy parameter is positive for all $r>r_0$ which gives the repulsive nature of geometry.

We have tried to find a WH solution within the framework of EGB theory by taking an specific shape function and radial dependent redshift function. We observed that with  an appropriate value of coupling constant $\alpha$ and wormhole throat $r_0$ one can obtain wormholes without the requirement of exotic matter. We have also analyzed solutions for $\alpha >0$ and $\alpha =0$, is consistent with previous results obtained in \cite{ref13}. 

%%%%%%%%%%%%%%%%%%%%%%%%%%%%%%%%%%%%%%%%%%%%%%%%%%%%%%%%%%%%%%%%%%%%%%%%%%%%%%%%%%%%%%%%%%%%%%%%%%%%%%5
%\section*{Acknowledgments} 

%%%%%%%%%%%%%%%%%%%%%%%%%%%%%%%%%%%%%%%%%%%%%%%%%%%%%%%%%%%%%%%%%%%%%%%%%%%%%%%%%%%%%%%%%%%%%%%%%%%%%%
%%%%%%%%%%%%%%%%%%%%%%%%%%%%%%%%%%%%%%%%%%%%%%%%%%%%%%%%%%%%%%%%%%%%%%%%%%%%%%%%%%%%%%%%%%%%%%%%%%%%%%%%%%%

\end{document}